\documentclass[acmlarge,nonacm]{acmart}

\AtBeginDocument{}

\usepackage{pdflscape}
\usepackage{afterpage}    

\usepackage{threeparttable}
\usepackage{tabularx}

\usepackage{longtable}
\usepackage{multirow}
\usepackage{colortbl}
\usepackage{adjustbox}
\usepackage{xspace}

\usepackage{wasysym}

\usepackage{multibib}
\newcites{apndx}{Additional References}

\newcommand{\rv}[1]{\rotatebox[origin=lB]{90}{\parbox{2.2cm}{\centering \textbf{#1}}}}
\newcommand{\thetitle}[0]{A Hitchhiker's Guide to Privacy-Preserving Digital Payment Systems: A Survey on Anonymity, Confidentiality, and Auditability}  
\newif\ifshowappendix
\showappendixtrue

\newif\ifarxiv
\arxivtrue

\newcommand{\refTabMixing}[0]{Appendix A, Table\ifshowappendix ~\ref{tab:sota-mixing}\else ~A.6\fi } 

\newcommand{\refAppWitnessEnc}[0]{\ifshowappendix Appendix~\ref{app:witenc}\else Appendix~A.4.5\fi } 

\newcommand{\refAppMixing}[0]{\ifshowappendix Appendix~\ref{apx:sec:mix:privacy}\else Appendix~A.3.2\fi } 

\newcommand{\tlong}[1]{\ifarxiv{\xspace#1\xspace}\fi} \newcommand{\cryptodef}[1]{\ifarxiv{#1}\fi}

\begin{document}

\title{\thetitle{}}
\renewcommand{\shorttitle}{A Hitchhiker's Guide to Privacy-Preserving Digital Payment
Systems}

\author{Matteo Nardelli}
\authornote{The views expressed in this paper are those of the authors and do not necessarily reflect the views of the Bank of Italy.}
\email{matteo.nardelli@bancaditalia.it}
\orcid{0000-0002-9519-9387}
\author{Francesco De Sclavis}
\authornotemark[1]
\email{francesco.desclavis@bancaditalia.it}
\orcid{0009-0004-1318-3878}
\author{Michela Iezzi}
\authornotemark[1]
\email{michela.iezzi@bancaditalia.it}
\orcid{0009-0008-1505-1843}
\affiliation{\institution{Bank of Italy}
  \city{Rome}
  \country{Italy}
}
\renewcommand{\shortauthors}{Nardelli et al.}

\begin{abstract}
Crypto-assets and central bank digital currencies (CBDCs) are reshaping 
how value is exchanged in distributed computing environments.
These systems combine cryptographic primitives, protocol design, and system architectures to provide transparency and efficiency while raising critical challenges around privacy and regulatory compliance.
This survey offers a comprehensive overview of privacy-preserving digital payment systems, covering both decentralized ledger systems and CBDCs. 
We present a 
taxonomy of privacy goals---including anonymity, confidentiality, unlinkability, and auditability---and map them to the cryptographic primitives, protocols, and system architectures that implement them.
Our work adopts a design-oriented perspective, linking high-level privacy objectives to concrete implementations. 
We also trace the evolution of privacy-preserving digital payment systems through three generations, highlighting shifts from basic anonymity guarantees toward more nuanced privacy-accountability trade-offs.
Finally, we identify open challenges, 
motivating further research into architectures and solutions that balance strong privacy with real-world auditability needs.
\end{abstract}

\maketitle

\section{Introduction}
\label{sec:intro}

The rapid proliferation of digital payment systems has sparked renewed interest in designing distributed architectures that balance user privacy with institutional accountability.
Spanning decentralized crypto-assets\footnote{The term \emph{cryptocurrency} is widely used to describe decentralized systems such as Bitcoin, Ethereum, or Monero, that enable peer-to-peer asset transfer without a central authority. However, this usage differs from the terminology adopted by central banks and regulators~\cite{ebc18bitcoin}, which reserve the term ``currency'' for instruments with legal tender status. To reflect this distinction and bridge the terminology gap between these communities, we adopt the more nuanced terms crypto-assets or digital payment systems throughout the paper.} and centrally managed infrastructures such as central bank digital currencies (CBDCs), these systems leverage cryptographic techniques and distributed consensus to enable secure, efficient, and programmable value transfer.
However, the same transparency and auditability that make these systems attractive also 
raise fundamental questions about user privacy, institutional control, and regulatory compliance (e.g.,~\cite{auer25privacy,ghesmati22:sokbitcoin,Herskind20}). 
For example, public blockchains typically offer pseudonymity.\footnote{Anonymity denotes indistinguishability within a set, while pseudonymity uses identifiers that are not directly linked to real-world identities.}
This enables transaction tracing, address clustering, and ultimately user de-anonymization, thus exposing 
financial behavior
(e.g.,~\cite{Meiklejohn13,moser18,Kappos18,Herskind20,Karame15}).
As a result, \emph{privacy-preserving digital payments} have emerged as a critical research field, attracting attention from cryptographers, engineers, economists, and policymakers.

Researchers have proposed a variety of privacy-preserving solutions leveraging privacy-enhancing technologies (PETs), including 
mixing protocols 
(e.g.,~\cite{Maxwell13:coinjoin,Ziegeldorf18coinparty,ruffing17valueshuffle}),  
confidential transactions (e.g.,~\cite{bunz20:zether,baudet22:zef,Guan22:blockmaze}), 
and privacy-native crypto-assets (e.g.,~\cite{Kiayias22:peredi,wust22:platypus,tomescu22:utt,noether16:monero}). 
These approaches provide different privacy guarantees, such as anonymity, confidentiality, and unlinkability.
However, privacy alone is not sufficient: regulatory frameworks such as Anti-Money Laundering (AML) and Combating the Financing of Terrorism (CFT) require mechanisms for auditability and selective disclosure. 
To reconcile these conflicting goals, recent work has explored cryptographic tools such as zero-knowledge proofs (ZKPs), threshold encryption, and homomorphic commitments.

This survey proposes a unifying, design-oriented perspective on privacy-preserving digital payment systems,
covering both decentralized ledgers and centrally managed systems with privacy guarantees---including CBDCs.
It organizes the literature around conceptual dimensions
that highlight recurring design choices and trade-offs.
Unlike previous surveys that focus on broad overviews (e.g.,~\cite{genkin18,auer25privacy,Herskind20}),
specific technologies (e.g.,~\cite{Almashaqbeh22,ghesmati22:sokbitcoin,sun21:surveyzkp}), 
or do not consider auditability (e.g.,~\cite{peng21:survey,Herskind20,feng19}), our work aims to bridge these dimensions.
Our contributions are threefold.
First, we introduce a taxonomy that classifies privacy-preserving digital payments systems along multiple dimensions, including system architecture, trust model, value representation, privacy guarantees, and auditability mechanisms.
Second, we systematize how privacy properties are formalized and realized, mapping high-level goals to technical mechanisms.
We trace the development of privacy-preserving digital payment systems, highlighting the role of cryptographic techniques across three generations.
Third, we identify open research challenges stemming from the design of privacy-preserving digital payment systems, which lie at the intersection of cryptography, distributed systems, and privacy engineering.
Supplementary material presents the methodology used to collect and select the surveyed literature, and further contextualizes key publication venues and technical details of specific privacy-enhancing mechanisms.

This paper is organized as follows.
Section~\ref{sec:related} reviews related surveys and outlines how our work complements and differs from them.
Section~\ref{sec:background} introduces the key cryptographic primitives used in privacy-preserving digital payments, which are then analyzed along key architectural and cryptographic dimensions in Section~\ref{sec:ppfeatures}.
Then, Section~\ref{sec:openchallenges} outlines open challenges and future research directions,
and Section~\ref{sec:conclusion} concludes the paper.
 \section{Related Surveys}
\label{sec:related}

A large body of literature has explored privacy-preserving digital payment infrastructures, driven by both the technical challenges and societal concerns about financial surveillance.

\paragraph{Privacy in Crypto-assets}
\tlong{The rapid growth of crypto-asset systems has spurred multiple efforts to systematize knowledge (SoK) across diverse technical communities.} 
Bonneau et al.~\cite{Bonneau15} reviewed extensions of Bitcoin and alternative protocols (sometimes referred to as altcoins), including early privacy-native protocols. 
Genkin et al.~\cite{genkin18} offered a broad and accessible overview of techniques for anonymity in decentralized crypto solutions, while emphasizing the absence of a unified definition of privacy.
More recent surveys attempted to formalize anonymity notions. For instance, Amarasinghe et al.~\cite{Amarasinghe19} developed a taxonomy of anonymity features\tlong{ (including anonymity, confidentiality, unlinkability, untraceability, and deniability)}, and evaluated Bitcoin-derived solutions against it. 
Our work draws from and extends this analysis by further considering the trade-offs between anonymity, confidentiality, and \emph{auditability} across a broader range of systems, including CBDCs.
Feng et al.~\cite{feng19} offered a layered classification of privacy-preserving mechanisms from network-level obfuscation to cryptographic primitives\tlong{ such as ring signatures, Pedersen commitments, homomorphic encryption, and NIZKs}.
While technically detailed, their coverage of payment protocols and privacy-accountability trade-offs is limited.
Similarly, a systematic literature review by Herskind et al.~\cite{Herskind20} catalogs techniques such as stealth addresses, confidential transactions, and network anonymity. However, it does not analyze core system features such as value representation models or system architectures\tlong{, which are key components of our taxonomy}.
Peng et al.~\cite{peng21:survey} identify PETs and illustrate how various protocols integrate them, but do not investigate protocol evolution and design trade-offs.

Other related surveys take a narrower or orthogonal perspective.
Almashaqbeh and Solomon~\cite{Almashaqbeh22} focus primarily on privacy in blockchain computation, i.e., smart contracts;
Zhang~\cite{zhang23} and Wang et al.~\cite{wang20:survey} 
overview privacy features in deployed public ledger-based systems;
Alsalami and Zhang~\cite{Alsalami19:sok} present 
an analysis of PETs, not considering accountability.

\paragraph{Privacy in CBDCs}
A distinct body of work addresses privacy in CBDCs, driven by 
the need to balance individual privacy with institutional oversight. Auer et al.~\cite{auer23privacy:short,auer25privacy} distinguish between {soft privacy}, based on access controls and trust in intermediaries, and {hard privacy}, based on cryptographic techniques. While they offer a high-level classification and conceptual framework, their discussion does not consider low-level protocol mechanisms.
\tlong{Other regulatory-focused contributions include} 
Darbha and Arora (Bank of Canada)\footnote{\url{https://www.bankofcanada.ca/2020/06/staff-analytical-note-2020-9/}} and Pocher and Veneris~\cite{Pocher22} explore PETs from a compliance and governance perspective, mapping privacy mechanisms to AML/CFT auditability requirements in retail CBDC designs.
These works offer blueprints but lack detailed cryptographic analysis.
Conversely, our work presents a comparative review of technical design choices including primitives, protocols, and architectures.

\paragraph{Specialized Surveys}
Several surveys focus on specific technologies (e.g., Bitcoin~\cite{ghesmati22:sokbitcoin}, smart contracts~\cite{Benarroch24:soksc,li22:soksc}, DeFi~\cite{baum23:sokdefi}, mixers~\cite{arbabi23:mixingsurvey}, accountability mechanisms~\cite{Chatzigiannis21:sokacc}, and ZKPs~\cite{sun21:surveyzkp,Almashaqbeh22,liang25:soksnarks}). These contributions do not consider the full design space of privacy-preserving payment systems as we do, nor do they explore how different privacy features interact---particularly in settings requiring auditability.
Koerhuis et al.~\cite{Koerhuis20} present a forensic analysis of Monero and Verge, and highlight how implementation flaws threaten privacy.

\paragraph{Contribution of This Survey}
We propose a unified taxonomy of privacy-preserving payment systems that integrates both 
decentralized ledger-based solutions and CBDCs. Specifically, we present a systematic analysis of existing solutions to identify key value representation models, privacy features, \tlong{i.e., anonymity, confidentiality, accountability, }and auditability guarantees.
We focus on how such features are realized through cryptographic protocols and specific design choices, resulting in a classification of contributions that 
connects design goals to technical implementations.
 \section{Background}
\label{sec:background}

We first introduce key cryptographic primitives (Sections~\ref{sec:pet:commitments}--\ref{sec:pet:zkp}), and then we briefly review key cryptographic protocols used in privacy-preserving digital payment systems (Section~\ref{sec:prot:sign}--\ref{sec:prot:otpk}).

\subsection{Commitments and Homomorphic Commitments} \label{sec:pet:commitments}
A commitment scheme is a cryptographic primitive that allows a party to commit to a value in such a way that it cannot be changed later.
The scheme consists of two phases: (i) \emph{commit phase}, where a commitment is made to a value $v$, but the value is kept secret from third parties; (ii) \emph{opening phase}, where the committed value $v$ is revealed. A commitment is \emph{hiding} if an adversary cannot distinguish which of two values corresponds to the commit; it is \emph{binding} if the committer cannot generate the same commit from two different values. These properties can be either \emph{computational}, if the adversary is computationally bounded, or \emph{perfect}, if not. 
A commitment scheme cannot be both perfectly hiding and perfectly binding, due to fundamental cryptographic limitations~\cite{goldreich04foundations}.

Homomorphic commitments preserve the addition or multiplication of corresponding values.
\cryptodef{For example, let $g$ and $h$ be elements of a prime $q$ group $\mathbb{G}_q$ such that $\log_gh$ is not feasible to compute, let $r_i$ be a random value over $\mathbb{Z}_q$ and let $v_i \in \mathbb{Z}_q$ be the committer's value. Perdersen commitments are defined as
$ c_{v_i}=g^{v_i}h^{r_i} $.
The homomorphic properties of Pedersen commitments guarantee that
$ g^{v_1}h^{r_1} \cdot g^{v_2}h^{r_2}=g^{v_1+v_2}h^{r_1+r_2} $
namely, the sum of the commitments of $v_1$ and $v_2$ is equal to the commitment of the sum of the two values under the sum of the two randomness $r_1$ and $r_2$, allowing multiple operations on the committed messages.}
In particular, Pedersen~\cite{Pedersen92}  commitments are perfectly hiding, computationally binding, and additively homomorphic; ElGamal~\cite{elgamal85} commitments are computationally hiding, perfectly binding, and multiplicatively homomorphic.

\subsection{Encryption Schemes}
\label{sec:pet:enc}

\paragraph{Identity-based encryption (IBE)} This scheme~\cite{shamir85:ibe} is a public-key encryption scheme in which the public key is a string that uniquely identifies a user (e.g., an email). The corresponding private keys for decryption are derived from a master private key, by a trusted central authority.
IBE can be used to simplify key management and recover, and access control within an organization. Anonymous variants make the public key indistinguishable from the ciphertext, effectively hiding the identity of the recipient of the message.

\paragraph{Threshold encryption} It allows data to be encrypted in such a way that it can only be decrypted when a predefined quorum of participants agrees to reveal it. Threshold encryption can help maintain privacy while removing single point of failures.
A popular implementation uses the ElGamal encryption scheme~\cite{elgamal85}.
A \emph{threshold-issuance} variant of Boneh-Franklin IBE is used in UTT~\cite{tomescu22:utt}, to distribute the master secret key among multiple authorities.

\paragraph{Homomorphic encryption} This technique~\cite{rivest1978} allows computing additions, multiplications or both, without revealing the plaintext. Operations can be performed directly on ciphertexts such that the result corresponds to the encryption of the same operation applied to the plaintexts\cryptodef{, i.e.,
$\mathrm{Enc}(m_1)\star\mathrm{Enc}(m_2)=\mathrm{Enc}(m_1\star m_2)$,
where $\mathrm{Enc}(\cdot)$ represent an encryption function and $\star$ denotes either addition or multiplication}.
This allows to delegate computation to a third party while maintaining confidentiality.
It enables {confidential transactions} when used as a building block for ZKPs, like to homomorphic commitments (e.g.,~\cite{Ma21:dsc,bunz20:zether}).

\subsection{Zero-knowledge Proofs} 
\label{sec:pet:zkp}

A ZKP is a protocol enabling 
a prover to convince a verifier 
that a statement is true, without disclosing any information beyond the validity of the statement itself. A dishonest prover can convince the verifier of a false statement, only with negligible probability (\emph{soundness}). If the prover is computationally bounded (\emph{computational soundness}), the proof is known as \emph{argument}~\cite{Bitansky12:zksnark}.
ZKPs can be either \emph{interactive}, requiring the parties to exchange messages, or non-interactive (NIZK)\tlong{,
when the verifier is convinced through a single message}.
ZKPs can also be \emph{proofs of knowledge} when the prover can convince the verifier that he knows a certain value, e.g., a secret key (or \emph{arguments of knowledge}, if the prover is computationally bounded).
If the ZKP is short and fast to verify, it is said to be \emph{succinct}. If it requires no trusted setup, it is \emph{transparent}.

A conceptual framework for ZK protocols is the $\Sigma$-protocol, which is a type of 3-move interactive proof system: the prover commits a value, the verifier provides a challenge, and the prover provides a proof based on the commit and the challenge, so that they cannot choose a ``malicious'' proof. Using the Fiat-Shamir transform, $\Sigma$-protocols can be made non-interactive~\cite{fiatshamir87}.
Building on this, various NIZK protocols have been designed, including zk-SNARK~\cite{Bitansky12:zksnark}, Bulletproofs~\cite{bunz18:bulletproof}, and---to some extent---zk-STARK~\cite{Ben-Sasson19:zkstark}.
Zk-SNARKs are succinct non-interactive arguments of knowledge. 
Many zk-SNARK constructions have been proposed in past years (e.g.,~\cite{Bitansky12:zksnark,boneh20:halo,groth16,chen23:reviewzksnarks}), with different approaches also for setup. 
Among them, the most popular approach is Groth16~\cite{groth16}, an instance optimized for performance but with some trade-offs like requiring a trusted setup. 
Bulletproofs are optimized for range proof and does not require trusted setup. 
Zk-STARK is a recent approach offering efficient proving times and no need for trusted setup. STARKs are succinct transparent arguments of knowledge. However, in the current state, zk-STARK generates large proof size (a few hundred kilobytes), so they are not as succinct as SNARKs. 
A recent survey on ZKPs can be found, e.g., in~\cite{Christ24:sok-rangeproofs,elhajii24:zkeval}.

\subsection{Anonymity-enhanced Signatures}
\label{sec:prot:sign}
\paragraph{Blind signatures} Blind signatures~\cite{chaum83} are a form of digital signature in which the signer produces a valid signature on a message without learning its content: before signing, the message is \emph{blinded}, i.e., transformed through a random element that makes it unrecognizable. Once the contents of the signed message are unblinded, they can be publicly verified as usual.

\paragraph{Randomizable signatures} They provide a way to randomize a previously generated signature, transforming it into a new valid signature, but making it impossible to link the two instances together.
This enables a level of anonymity by unlinking user identities from their transaction history, as each transaction appears distinct even if performed by the same user.
The original idea was proposed by Camenisch and Lysyanskaya~\cite{Camenisch-Lysyanskaya03, Camenisch-Lysyanskaya04}\tlong{ but suffered from a linear size in the number of messages to be signed}. Later, Pointcheval-Sanders proposed more efficient signature and verification algorithms~\cite{pointcheval16:randsig}.
Existing works (e.g.,~\cite{baudet22:zef,wust19:prcash,Sarencheh25:parscoin, tomescu22:utt}) adopt either Pointcheval-Sanders or Coconut~\cite{sonnino20:coconut}, a threshold scheme derived from it.
The Pointcheval-Sanders scheme can produce signatures of a message directly from its commitment; this effectively makes it a blind signature, thanks to the hiding property of commitments.

\paragraph{Group signatures} This scheme~\cite{Chaum-vanHeyst91:groupsig} enables 
any member of a designated group to generate a valid signature without revealing who signed (ensuring anonymity).
A verifier uses a {group public key} to confirm that a valid signature was produced by some group member.
This property also implies \emph{unlinkability}: it is not possible to determine whether two different signatures were produced by the same user~\cite{Perera22:survey-GSRS}. However, group signatures can be made \emph{linkable} to prevent double spending in transactions (e.g.,~\cite{Zhang19}). 
Group signatures also support \emph{user traceability}: a designated group manager holds a {group secret key} that enables him to revoke a signer's anonymity and reveal his identity when necessary.

\paragraph{Ring signatures} Ring signatures~\cite{rivest01:ringsig} are similar to group signatures, as they hide the signer's identity behind a group. Differently from group signatures, user anonymity is permanent, because there is no revoke mechanism. Indeed, the signer simply joins their public key with a set of other public keys during the signing phase. 
A verifier needs the whole list of public keys to verify the signature validity, and cannot discern which \cryptodef{of the $n$} public key\cryptodef{s} is of the signer, but can only guess \cryptodef{with probability $1/n$}.
There are some variant ring signatures, where anonymity is somewhat limited and the signer can be traced under special circumstances, e.g., after double spending. 
These are called \emph{traceable} ring signatures ~\cite{fujisaki2007,fujisaki2011}. 
Ring signatures can also be \emph{linkable}~\cite{liu:linkablering}, 
if they allow anyone to determine if two signatures were produced by the same signer (e.g.,~\cite{vanSaberhagen13cryptonote,li21:traceable-monero}).

\paragraph{Threshold signatures} They are a type of signature that can be jointly produced by a committee of signers. A $t$-of-$n$ signature scheme requires at least $t$ signers out of a total of $n$ participants to produce a valid signature.
Works that use threshold signatures~\cite{Androulaki20, Sarencheh25:parscoin, Kiayias22:peredi, baudet22:zef} use threshold blind signatures, based on Coconut~\cite{sonnino20:coconut}.

\subsection{Secure Multi-party Computation}
\label{sec:prot:mpc}

Secure Multi-Party Computation (MPC) allows multiple parties to jointly compute a result over their private inputs without any party having to reveal its own input to the others.
The key idea is that the parties collaborate to perform some computation and, at the end, they all learn the correct output of the function, but nothing else.
Beyond applications such as CoinParty~\cite{Ziegeldorf18coinparty}, MPC is also used to decentralize trusted setup procedures in zk-SNARKs, enabling
security to hold as long as at least one participant behaves honestly.

\subsection{One-time Public Keys}
\label{sec:prot:otpk}
\paragraph{Stealth addresses} They hide the recipient's identity making each payment unlinkable. They are one-time public keys generated by the transaction sender with public key provided by the receiver.
Recipient does not update his public key, but the sender can derive fresh one-time keys for each transaction. 
Let $\mathrm{G}$ be a generator of an elliptic curve group $\mathbb{G}$ with scalar field $\mathbb{Z}_q$. Each receiver $ R $ has a public view key $V=vG$ and a public spend key $S=sG$, where $v,s\in \mathbb{Z}_q$ are secret scalars. 
A sender chooses a random scalar $r\in \mathbb{Z}_q$ and computes the shared secret $H=\mathcal{H}_p(rV) \in \mathbb{Z}_q$, with $\mathcal{H}_p(\cdot) $ a hash-to-scalar function, and the one-time public key $P=HG+S$. The sender includes $ P $ and $R=rG$ in the transaction. The recipient can scan the ledger and for each $ P $ and $R$ computes $P^\prime = \mathcal{H}_p(vR)G + S $. If $P^\prime = P$ then the output belongs to the recipient.

\paragraph{Updatable public keys} They allow to update the recipient's public key in a decentralized manner.
Quisquis~\cite{Fauzi19:quisquis} introduces this technique to update both sender and recipient public keys. 
Given a prime-order group $\mathbb{G}_p$ with $p$ the prime order of the group, and $ G $ its generator, each user has a secret key $s_i$ and a public key $ P_i = (G^r, G^{r \cdot s_i}) = (U,V) $, for some random value $r \in \mathbb{Z}_q$. 
To perform a transaction, the sender updates the public key by choosing a random value $ z \in \mathbb{Z}_q$ and computing $P^\prime_i = (U^{z}, V^{z}) $. The updated key is indistinguishable from a fresh key\cryptodef{, thanks to the Decisional Diffie-Hellman assumption}.
The recipient 
verifies whether a given updated public key $P^\prime_i$ corresponds to his secret key $s_i$. This works because updated keys still verify correctly with the original secret key.
 \begin{figure}[t]
    \centering
    \includegraphics[width=.87\linewidth]{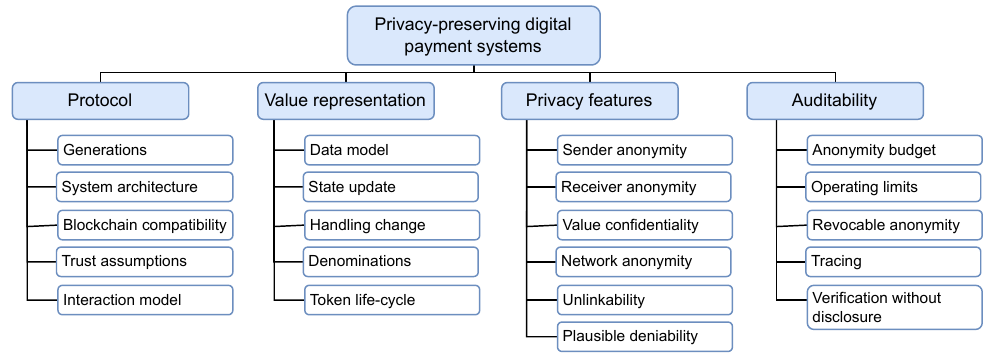}
    \caption{Dimension of analysis along which we review privacy-preserving digital payment systems.}
    \label{fig:taxonomy}
    \Description{Dimension of analysis along which we review privacy-preserving digital payment systems.}
\end{figure}

\begin{table}
{
    \renewcommand{\arraystretch}{0.9}
    \setlength{\tabcolsep}{3pt}
    \centering
    \caption{Labels and notation used in Table~\ref{tab:sota-cryptocurrencies}.}
    \label{tab:sota_legend}
    \adjustbox{max width=\columnwidth}{\footnotesize
    \begin{tabular}{lrlrl}
        \toprule
        \textbf{Dimension} & \multicolumn{4}{l}{\textbf{Legend}}\\
        \midrule
        \multicolumn{3}{l}{\emph{Protocol}}\\
        {Type } & CD & CBDC & D & Digital payment system \\
        {Blockchain Compatibility} & B & Bitcoin & E & Ethereum \\
\midrule
        \multicolumn{3}{l}{\emph{Value Representation}}\\
        {Data Model} & A & Account-based & CT & Chaumian token \\
        {} & T2B & Transaction per bank table & UT & Unspent token \\
\midrule
        \multicolumn{3}{l}{\emph{Other}}\\
        {Auditability} & AB & Anonymity budget & L & Operating limits \\
        { } & RA & Revocable anonymity & TR & Coin tracing \\
        { } & ZD & \multicolumn{3}{l}{Verification without disclosure} \\
        \cmidrule(r){2-5}
        {Cryptographic Primitives} & BS & Blind signature & C & Commitment \\
         & DS & Digital signature & GS & Group signatures \\
        {} & HC & Homomorphic commitment & IBE & Identity-based encryption \\
        {} & MS & Multi signature & MT & Merkle tree \\
        {} & O-RAM & Oblivious RAM & PE & Public key encryption \\
        {} & RS & Ring signature & SA & Stealth address \\
        {} & UP & Updatable public keys & ZK & Zero-knowledge proof \\
\cmidrule(r){2-5}
        {} & $X^t$ & Threshold-based version of $X$ 
           & $X^r$ & Randomizable version of $X$ \\
        {} & $X$* & Custom variant of $X$ & & \\ 
        \midrule
        {\emph{Notation}} & \checkmark & Yes & $\times$ & No\\
        {} & \LEFTcircle & Partial & $\Leftrightarrow$ & Alternative \\
        {} & \Circle & Abstracted & & \\
        \bottomrule
    \end{tabular}
    }
    \renewcommand{\arraystretch}{1.0}
}
\end{table}

\section{Design Dimensions of Privacy-Preserving Digital Payment Systems}
\label{sec:ppfeatures}

The research community has explored various solutions to ensure anonymity, confidentiality, and auditability in crypto-assets and digital payment systems.  
We identify key dimensions of analysis---protocol, value representation,
privacy features, 
and auditability---and highlight the most significant design choices.\footnote{
To avoid terminological ambiguity, we align with De Roode and Everts~\cite{deRoode25}, who distinguish privacy (a user right) from privacy-preservation (a system property) and define anonymity, unlinkability, and auditability in a unified DLT terminology.}
These dimensions are interconnected and jointly embed privacy features into the system.
Figure~\ref{fig:taxonomy} outlines the dimensions that structure this section.
Table~\ref{tab:sota_legend} describes the labels used in Table~\ref{tab:sota-cryptocurrencies}, which summarizes our literature review.\footnote{Some properties shown in Figure~\ref{fig:taxonomy} and Table~\ref{tab:sota-cryptocurrencies} (e.g., double-spending protection, ledger state updates) are treated as supporting mechanisms 
underpinning correctness, privacy, or auditability, and are discussed within those dimensions rather than as independent classification axes.}

\subsection{Protocol}
\label{sec:ppf:prot}
The \emph{protocol} dimension outlines the main system design choices through which existing solutions can be compared.

\afterpage{
\begin{landscape}
\renewcommand{\arraystretch}{0.9}
\setlength{\tabcolsep}{2.3pt}
\footnotesize
\begin{longtable}{lcccccccccccccccccccl}
\caption{Classification of existing approaches for privacy-preserving digital payments and crypto-asset exchanges (see notation in Table~\ref{tab:sota_legend}).}
\label{tab:sota-cryptocurrencies} \\
\toprule
\rowcolor{white}
\multicolumn{1}{l}{\textit{}} 
        & \multicolumn{7}{c}{\textit{Protocol}} 
        & \multicolumn{5}{c}{\textit{Value Representation}} 
        & \multicolumn{6}{c}{\textit{Privacy}} 
        & \multicolumn{2}{c}{\textit{Other}} 
        \\
\cmidrule(r){2-8}
\cmidrule(r){9-13}
\cmidrule(r){14-19}
\cmidrule(r){20-21}
\rowcolor{white}
\textbf{Paper} 
        & \rv{Year} 
            & \rv{Type} 
            & \rv{Decentralized} 
            & \rv{Blockchain Compatibility}
            & \rv{No-Trust Assumptions}
            & \rv{Non-Interactive} 
            & \rv{Implemented}
        & \rv{Data Model}
            & \rv{Change} 
            & \rv{Different Denominations} 
            & \rv{Transferable Token} 
            & \rv{Immediate Claiming} 
        & \rv{Sender Anonymity} 
            & \rv{Receiver Anonymity}
            & \rv{Transaction Value Confidentiality}
            & \rv{Network Anonymity} 
            & \rv{Unlinkability} 
            & \rv{Plausible Deniability} 
        & \rv{Auditability}
            & \rv{Cryptographic Primitives} 
    \\
\midrule
\endfirsthead
    \multicolumn{21}{c}{\textit{(\tablename~\thetable{} continued from previous page)}} \\
\multicolumn{1}{l}{\textit{}} 
        & \multicolumn{7}{c}{\textit{Protocol}} 
        & \multicolumn{5}{c}{\textit{Value Representation}} 
        & \multicolumn{6}{c}{\textit{Privacy}} 
        & \multicolumn{2}{c}{\textit{Other}} 
        \\
\cmidrule(r){2-8}
\cmidrule(r){9-13}
\cmidrule(r){14-19}
\cmidrule(r){20-21}
\rowcolor{white}
\textbf{Paper} 
        & \rv{Year} 
            & \rv{Type} 
            & \rv{Decentralized} 
            & \rv{Blockchain Compatibility}
            & \rv{No-Trust Assumptions}
            & \rv{Non-Interactive} 
            & \rv{Implemented}
        & \rv{Data Model}
            & \rv{Change} 
            & \rv{Different Denominations} 
            & \rv{Transferable Token} 
            & \rv{Immediate Claiming} 
        & \rv{Sender Anonymity} 
            & \rv{Receiver Anonymity}
            & \rv{Transaction Value Confidentiality}
            & \rv{Network Anonymity} 
            & \rv{Unlinkability} 
            & \rv{Plausible Deniability} 
        & \rv{Auditability}
            & \rv{Cryptographic Primitives} 
    \\
\midrule
\endhead
\multicolumn{21}{c}{\textit{(continued on next page)}}\\\endfoot
\endlastfoot
Chaum~\cite{chaum83} & 1983 & D & $\times$ & $\times$ & \checkmark & $\times$ & $\times$ & CT & $\times$ & $\times$ & $\times$ & \checkmark & \checkmark & $\times$ & $\times$ & $\times$ & $\times$ & \checkmark & $\times$ & BS  \\ 
\rowcolor{gray!10}
Hayes~\cite{hayes90} & 1990 & D & $\times$ & $\times$ & $\times$ & $\times$ & $\times$ & CT & $\times$ & $\times$ & \checkmark & $\times$ & \LEFTcircle & $\times$ & $\times$ & $\times$ & $\times$ & \checkmark & $\times$ & BS*  \\ 
Sander and Ta-Shma~\cite{Sander99} & 1999 & D & $\times$ & $\times$ & $\times$ & $\times$ & $\times$ & CT & $\times$ & $\times$ & $\times$ & $\times$ & \LEFTcircle & $\times$ & $\times$ & $\times$ & $\times$ & \checkmark & RA & MT  \\ 
\rowcolor{gray!10}
Camenisch et al.~\cite{Camenisch06} & 2006 & D & $\times$ & $\times$ & $\times$ & $\times$ & $\times$ & CT & $\times$ & $\times$ & $\times$ & $\times$ & \checkmark & $\times$ & \checkmark & $\times$ & $\times$ & \checkmark & AB & HC, PE*, ZK  \\ 
CryptoNote~\cite{vanSaberhagen13cryptonote} & 2013 & D & \checkmark & $\Leftrightarrow$  & \checkmark & \checkmark & $\times$ & UT & \checkmark & \checkmark & $\times$ & \checkmark & \checkmark & \checkmark & $\times$ & $\times$ & \checkmark & $\times$ & $\times$ & RS, SA  \\ 
\rowcolor{gray!10}
ZeroCoin~\cite{miers13zerocoin} & 2013 & D & \checkmark & \LEFTcircle & $\times$ & \checkmark & $\times$ & UT & \checkmark & $\times$ & $\times$ & \checkmark & \checkmark & $\times$ & $\times$ & $\times$ & \checkmark & $\times$ & $\times$ & HC, ZK  \\ 
EZC~\cite{Androulaki14} & 2014 & D & \checkmark & \LEFTcircle & \checkmark & \checkmark & $\times$ & UT & \checkmark & \checkmark & $\times$ & \checkmark & \checkmark & \LEFTcircle & \checkmark & $\times$ & \checkmark & $\times$ & $\times$ & HC, ZK  \\ 
\rowcolor{gray!10}
Monero~\cite{noether16:monero} & 2014 & D & \checkmark & $\Leftrightarrow$  & \checkmark & \checkmark & \checkmark\footnote{\url{https://github.com/monero-project/monero}}& UT & \checkmark & \checkmark & $\times$ & \checkmark & \checkmark & \checkmark & \checkmark & \checkmark & \checkmark & \checkmark & $\times$ & PE, RS, SA   \\ 
ZeroCash~\cite{BenSasson14:zerocash} & 2014 & D & \checkmark & \LEFTcircle & $\times$ & \checkmark & $\times$ & UT & \checkmark & \checkmark & $\times$ & \checkmark & \checkmark & \checkmark & \checkmark & $\times$ & \checkmark & $\times$ & $\times$ & C, PE, ZK  \\ 
\rowcolor{gray!10}
MimbleWimble~\cite{poelstra16:mimblewimble} & 2016 & D & \checkmark & $\Leftrightarrow$  & \LEFTcircle & $\times$ & $\times$ & UT & \checkmark & \checkmark & $\times$ & \checkmark & \checkmark & \checkmark & \checkmark & $\times$ & \LEFTcircle & \checkmark & $\times$ & HC, MS, ZK  \\ 
Garman et al.~\cite{Garman17} & 2017 & D & \checkmark & \LEFTcircle & $\times$ & \checkmark & $\times$ & UT & \checkmark & \checkmark & $\times$ & \checkmark & \checkmark & \checkmark & \checkmark & $\times$ & \checkmark & $\times$ & L,RA,TR & DS, PE, ZK  \\ 
\rowcolor{gray!10}
GNU Taler~\cite{dold19:taler} & 2017 & D & $\times$ & $\times$ & $\times$ & $\times$ & $\times$ & CT & \checkmark & \checkmark & $\times$ & \checkmark & \checkmark & $\times$ & $\times$ & $\times$ & \checkmark & \checkmark & $\times$ & BS  \\ 
Solidus~\cite{Cecchetti17} & 2017 & CD & $\times$ & \Circle & $\times$ & $\times$ & \checkmark & A & \checkmark & \checkmark & $\times$ & \checkmark & \checkmark & \checkmark & \checkmark & \Circle & \checkmark & \checkmark & RA, ZD & PE, O-RAM, ZK  \\ 
\rowcolor{gray!10}
zkLedger~\cite{Narula18} & 2018 & CD & \checkmark & $\Leftrightarrow$  & \checkmark & \checkmark & \checkmark & T2B & \checkmark & \checkmark & $\times$ & \checkmark & \checkmark & \checkmark & \checkmark & $\times$ & \checkmark & \checkmark & ZD & HC, PE  \\ 
PRCash~\cite{wust19:prcash} & 2019 & D & \LEFTcircle & $\Leftrightarrow$  & $\times$ & $\times$ & \checkmark & UT & \checkmark & \checkmark & $\times$ & \checkmark & \checkmark & \checkmark & \checkmark & $\times$ & \LEFTcircle & \checkmark & AB & DS$^r$, HC, ZK  \\ 
\rowcolor{gray!10}
Quisquis~\cite{Fauzi19:quisquis} & 2019 & D & \checkmark & $\Leftrightarrow$  & \checkmark & \checkmark & $\times$ & A & \checkmark & \checkmark & $\times$ & \checkmark & \checkmark & \checkmark & \checkmark & $\times$ & \checkmark & \checkmark & $\times$ & HC, UP, ZK  \\ 
Traceable Monero~\cite{li21:traceable-monero} & 2019 & D & \checkmark & $\times$ & \checkmark & \checkmark & \checkmark & UT & \checkmark & \checkmark & $\times$ & \checkmark & \checkmark & \checkmark & \checkmark & \checkmark & \checkmark & \checkmark & RA & PE, RS, SA  \\ 
\rowcolor{gray!10}
Zhang et al.~\cite{Zhang19} & 2019 & CD & $\times$ & $\Leftrightarrow$  & $\times$ & \checkmark & \checkmark & A & \checkmark & \checkmark & $\times$ & \checkmark & \LEFTcircle & \LEFTcircle & $\times$ & $\times$ & \LEFTcircle & \checkmark & RA & GS  \\ 
Androulaki et al.~\cite{Androulaki20} & 2020 & D & \checkmark & $\Leftrightarrow$  & $\times$ & $\times$ & $\times$ & UT & \checkmark & \checkmark & $\times$ & \checkmark & \LEFTcircle & \LEFTcircle & \LEFTcircle & $\times$ & \LEFTcircle & \checkmark & RA & BS$^t$, HC, PE, ZK  \\ 
\rowcolor{gray!10}
Barki and Gouget~\cite{barki20} & 2020 & D & \checkmark & $\times$ & $\times$ & $\times$ & $\times$ & CT & \checkmark & \checkmark & \checkmark & $\times$ & \LEFTcircle & \LEFTcircle & $\times$ & $\times$ & \LEFTcircle & $\times$ & RA, TR & BS, PE  \\ 
DCAP~\cite{lin20} & 2020 & D & \checkmark & $\Leftrightarrow$  & $\times$ & $\times$ & \LEFTcircle\footnote{\url{https://github.com/colyn91/-Smart-Contract-on-DCAP}}& A & \checkmark & \checkmark & $\times$ & \checkmark & \LEFTcircle & \LEFTcircle & $\times$ & $\times$ & \LEFTcircle & $\times$ & RA & ZK  \\ 
\rowcolor{gray!10}
PGC~\cite{chen20:pgc} & 2020 & D & \checkmark & $\Leftrightarrow$  & $\times$ & \checkmark & \checkmark\footnote{\url{https://github.com/yuchen1024/Kunlun/tree/master/adcp}}& A & \checkmark & \checkmark & $\times$ & \checkmark & $\times$ & $\times$ & \LEFTcircle & $\times$ & $\times$ & \checkmark & RA & DS, PE, ZK  \\ 
Zether~\cite{bunz20:zether} & 2020 & D & $\times$ & \checkmark~~(E) & \checkmark & \checkmark & \checkmark & A & \checkmark & \checkmark & $\times$ & \checkmark & \checkmark & \checkmark & \checkmark & $\times$ & \checkmark & $\times$ & $\times$ & HE, ZK  \\ 
\rowcolor{gray!10}
DSC~\cite{Ma21:dsc} & 2021 & D & \checkmark & $\Leftrightarrow$  & \checkmark & \checkmark & $\times$ & A & \checkmark & \checkmark & $\times$ & \checkmark & $\times$ & $\times$ & \checkmark & $\times$ & $\times$ & $\times$ & $\times$ & HE, ZK  \\ 
Gross et al.~\cite{gross2021} & 2021 & CD & \checkmark & $\times$ & $\times$ & $\times$ & $\times$ & A & \checkmark & \checkmark & $\times$ & \checkmark & \checkmark & \checkmark & \checkmark & $\times$ & \checkmark & $\times$ & L & C, ZK  \\ 
\rowcolor{gray!10}
MiniLedger~\cite{Chatzigiannis21} & 2021 & CD & \checkmark & $\Leftrightarrow$  & \LEFTcircle & \checkmark & \checkmark\footnote{\url{https://github.com/PanosChtz/Miniledger}}& T2B & \checkmark & \checkmark & $\times$ & \checkmark & \checkmark & \checkmark & \checkmark & $\times$ & \checkmark & \checkmark & ZD & HC, MT, PE, ZK  \\ 
PPChain~\cite{Lin21} & 2021 & D & \checkmark & $\Leftrightarrow$  & $\times$ & \checkmark & \checkmark & A & \checkmark & \checkmark & $\times$ & \checkmark & \LEFTcircle & \LEFTcircle & \LEFTcircle & $\times$ & $\times$ & $\times$ & RA & GS, PE  \\ 
\rowcolor{gray!10}
AT-CBDC~\cite{liu22} & 2022 & CD & \checkmark & $\Leftrightarrow$  & $\times$ & $\times$ & \checkmark & A & \checkmark & \checkmark & $\times$ & \checkmark & \checkmark & \checkmark & $\times$ & $\times$ & $\times$ & \checkmark & RA & BS, ZK  \\ 
BlockMaze~\cite{Guan22:blockmaze} & 2022 & D & \checkmark & $\Leftrightarrow$  & \checkmark & $\times$ & \checkmark\footnote{\url{https://github.com/Agzs/BlockMaze}}& A & \checkmark & \checkmark & $\times$ & \checkmark & \checkmark & \checkmark & \checkmark & $\times$ & \checkmark & $\times$ & $\times$ & C, ZK  \\ 
\rowcolor{gray!10}
Bontekoe et al.~\cite{bontekoe22} & 2022 & D & \LEFTcircle & \LEFTcircle & $\times$ & \checkmark & \checkmark\footnote{\url{https://github.com/TariqTNO/anonymous-transactions}}& UT & \checkmark & \checkmark & $\times$ & \checkmark & \checkmark & \checkmark & \checkmark & $\times$ & \checkmark & \checkmark & AB, RA & C, MT, ZK  \\ 
ERCE~\cite{xue23} & 2022 & D & \checkmark & $\Leftrightarrow$  & $\times$ & \checkmark & \LEFTcircle & UT & \checkmark & \checkmark & $\times$ & \checkmark & \LEFTcircle & \LEFTcircle & \checkmark & $\times$ & \LEFTcircle & $\times$ & L,RA & C, MT, ZK  \\ 
\rowcolor{gray!10}
PEReDi~\cite{Kiayias22:peredi} & 2022 & CD & \checkmark & $\times$ & $\times$ & $\times$ & $\times$ & A & \checkmark & \checkmark & $\times$ & \checkmark & \checkmark & \checkmark & \checkmark & $\times$ & \checkmark & \checkmark & RA & BS$^t$, C, PE$^t$, ZK  \\ 
Platypus~\cite{wust22:platypus} & 2022 & CD & $\times$ & $\times$ & \checkmark & $\times$ & \checkmark & A & \checkmark & \checkmark & $\times$ & \checkmark & \checkmark & \checkmark & \checkmark & \Circle & \checkmark & \checkmark & L, ZD & C, DS, ZK  \\ 
\rowcolor{gray!10}
UTT~\cite{tomescu22:utt} & 2022 & CD & \checkmark & $\times$ & $\times$\footnote{Threshold encryption relaxes trust requirements from a single entity to a quorum of entities.} & \checkmark & \checkmark\footnote{\url{https://github.com/definitelyNotFBI/utt}}& CT & \checkmark & \checkmark & $\times$ & \checkmark & \checkmark & \checkmark & \checkmark & \Circle & \checkmark & \checkmark & AB & DS$^{t,r}$, HC, IBE$^t$, ZK  \\ 
Zef~\cite{baudet22:zef} & 2022 & D & \checkmark & \checkmark~~(E) & \checkmark & $\times$ & \checkmark\footnote{\url{https://github.com/novifinancial/fastpay/tree/extensions}}& A & \checkmark & \checkmark & $\times$ & \checkmark & \checkmark & \checkmark & \checkmark & \Circle & \checkmark & $\times$ & $\times$ & BS$^r$, C, DS$^t$, SA, ZK  \\ 
\rowcolor{gray!10}
Bank of Australia~\cite{australia23} & 2023 & CD & $\times$ & \checkmark~~(E) & $\times$ & \checkmark & $\times$ & A & \checkmark & \checkmark & \checkmark & \checkmark & $\times$ & $\times$ & \checkmark & $\times$ & $\times$ & \checkmark & $\times$ & PE  \\ 
BIS Tourbillon~\cite{bis23:tourbillon} & 2023 & CD & $\times$ & $\times$ & \checkmark & $\times$ & $\times$ & CT/UT & $\times$ & \checkmark\footnote{Denominations in power of 2} & $\times$ & \checkmark & \checkmark & $\times$ & $\times$ & $\times$ & \LEFTcircle & \checkmark & $\times$ & BS  \\ 
\rowcolor{gray!10}
Goodell et al.~\cite{Goodell23} & 2023 & CD & \checkmark & $\Leftrightarrow$  & \checkmark & $\times$ & $\times$ & CT & \LEFTcircle & $\times$ & \checkmark & $\times$ & \checkmark & $\times$ & $\times$ & $\times$ & $\times$ & \checkmark & $\times$ & BS, PE  \\ 
KAIME~\cite{Dogan23:kaime} & 2023 & CD & \checkmark & $\times$ & \checkmark & \checkmark & \checkmark\footnote{\url{https://github.com/midmotor/kaime_cbdc_proof_test}}& A & \checkmark & \checkmark & $\times$ & \checkmark & \LEFTcircle & \LEFTcircle & \checkmark & $\times$ & $\times$ & $\times$ & RA & HC,PE$^t$, ZK  \\ 
\rowcolor{gray!10}
Lin et al.~\cite{lin23} & 2023 & D & \checkmark & $\Leftrightarrow$  & $\times$ & \checkmark & \checkmark & UT & \checkmark & \checkmark & $\times$ & \checkmark & \LEFTcircle & \LEFTcircle & \checkmark & $\times$ & \checkmark & $\times$ & RA & HE, PE, SA  \\ 
AQQUA~\cite{Papadoulis24:aqqua} & 2024 & D & \LEFTcircle & $\Leftrightarrow$  & $\times$ & \checkmark & $\times$ & A & \checkmark & \checkmark & $\times$ & \checkmark & \checkmark & \checkmark & \checkmark & $\times$ & \checkmark & \checkmark & L & HC, UP, ZK  \\ 
\rowcolor{gray!10}
PayOff~\cite{beer24:payoff} & 2024 & CD & $\times$ & $\times$ & $\times$ & \checkmark & \checkmark & A & \checkmark & \checkmark & \checkmark & $\times$ & \checkmark & \checkmark & \checkmark & $\times$ & \LEFTcircle & \checkmark & RA & C, ZK  \\ 
PCRAP~\cite{elfadul24} & 2024 & D & \checkmark & $\Leftrightarrow$  & $\times$ & $\times$ & $\times$ & UT & \checkmark & \checkmark & $\times$ & \checkmark & \checkmark & \checkmark & \checkmark & $\times$ & \checkmark & \checkmark & $\times$ & HC, MT, RS, SA  \\ 
\rowcolor{gray!10}
PICTURE~\cite{Zhang24} & 2024 & D & \checkmark & $\Leftrightarrow$  & $\times$ & \checkmark & \checkmark & UT & \checkmark & \checkmark & $\times$ & \checkmark & \checkmark & \checkmark & \checkmark & \Circle & \checkmark & \checkmark & RA & HC,PE$^t$, RS, SA, ZK  \\ 
Yu et al.~\cite{Yu24} & 2024 & D & $\times$ & \LEFTcircle & $\times$ & \checkmark & $\times$ & UT & $\times$ & $\times$ & $\times$ & \checkmark & \LEFTcircle & $\times$ & $\times$ & $\times$ & \LEFTcircle & $\times$ & RA & HC, PE  \\ 
\rowcolor{gray!10}
Friolo et al.~\cite{friolo25} & 2025 & D & \checkmark & $\times$ & $\times$ & $\times$ & $\times$ & CT & $\times$ & $\times$ & $\times$ & \checkmark & \checkmark & \checkmark & \checkmark & $\times$ & \checkmark & \checkmark & $\times$ & PE, ZK  \\ 
PARScoin~\cite{Sarencheh25:parscoin} & 2025 & D & \checkmark & \checkmark~~(E) & \checkmark & \checkmark & $\times$ & A & \checkmark & \checkmark & $\times$ & $\times$ & \checkmark & \checkmark & \checkmark & $\times$ & $\times$ & \checkmark & $\times$ & BS$^{t,r}$, HC, PE  \\ 
\bottomrule
\end{longtable}
\renewcommand{\arraystretch}{1.0}
\end{landscape}
} 

\subsubsection{Generations of Privacy-preserving Digital Payment Systems}

We introduce the notion of \emph{generations} of privacy-preserving digital payment systems to group works that share similar assumptions, deployment contexts, and design goals. 
Under this lens, first-generation systems typically rely on centralized issuers and focus on payer anonymity;
second-generation systems emphasize decentralized value transfer, unlinkability, and untraceability; 
while third-generation systems balance privacy with auditability and regulatory compliance.\footnote{These groupings are not strictly chronological, and several systems naturally exhibit hybrid characteristics across generations. The ``generations'' serve as conceptual lens to interpret recurring design goals and
assumptions across the dimensions discussed in this section.}

\paragraph{First Generation}
The \emph{first generation} of privacy-preserving digital payment systems focuses on centralized value transfer architectures, where a trusted issuing authority (typically a bank) is responsible for minting and redeeming digital coins, while cryptographic mechanisms are used to protect the payer's anonymity during spending. The central challenge in this setting is reconciling the need for user anonymity with the requirement that the issuer validate, redeem, and potentially audit issued tokens.
Chaum~\cite{chaum83} introduces the seminal untraceable payment system, known as \emph{ecash}\footnote{Later, Chaum et al.~\cite{chaum21} show how these mechanisms can be embedded into a two-tier, regulator-compliant retail CBDC architecture that preserves cash-like payer privacy while enforcing income transparency and AML/CFT through commercial banks.}, which employs blind signatures to prevent the issuing bank from linking withdrawn coins to subsequent payments (see Section~\ref{sec:ppf:value} for details).
Building on this foundation, a large body of work explored extensions 
of ecash~\cite{Andola21:survey-ecash}, including 
different coin denominations (e.g.,~\cite{dold19:taler}), 
partial spending (e.g.,~\cite{dold19:taler,OkamotoOhta91}), 
token transferability (e.g.,~\cite{hayes90,sai00,Baldimtsi15tr,Bauer21,Chase14}),
offline spending (e.g.,~\cite{chaum90}),  
and controlled forms of auditability (e.g.,~\cite{Camenisch06,Schoenmakers1998,Brands94,CamenischLysyanskaya01,Batten19}).\footnote{In contrast to Chaum's ecash, Netcoin~\cite{tewari-nuallain15} proposes a fiat-backed mint-based system in which coins are bound to user identities and fully traceable (no anonymity). This highlights how privacy was not a universal requirement in early centralized ecash designs.}

\paragraph{Second Generation}
The \emph{second generation} emerged with the introduction of Bitcoin~\cite{nakamoto08bitcoin}.
Bitcoin and, later, Ethereum~\cite{buterin13ethereum} propose public and permissionless blockchains, meaning that the ledger contains transactions visible to everyone and anyone can participate in the process of updating the public ledger.
Being public, these blockchains use addresses
to exchange assets in a pseudo-anonymous manner, as addresses are not directly tied to user identities.\footnote{Bitcoin addresses are designed to be ephemeral (though often reused), whereas Ethereum relies on persistent pseudonymous accounts.}
Public blockchains cannot easily and fully satisfy two privacy-related features: untraceability and unlinkability~\cite{rahalkar24}. 
\emph{Untraceability} ensures that, for any given incoming transaction, all potential senders appear equally likely. 
\emph{Unlinkability} means that, for any two outgoing transactions, there is no way to prove they were sent to the same recipient.
Untraceability and unlinkability can arise at different layers:
at the ledger level, they refer to preventing correlations between on-chain records;
at the network level, they refer to preventing correlations from communication patterns\footnote{Unless otherwise stated, we primarily focus on ledger-level privacy, and discuss network-layer aspects in Section~\ref{sec:ppf:net}.}.
While generating a new address for each transaction can enhance these properties, various studies indicate that achieving true anonymity and unlinkability on public blockchains remains challenging. An analysis of the transaction graph can reveal clusters of pseudonyms or patterns that link with high probability two or more addresses (e.g.,~\cite{Meiklejohn13,moser18,Kappos18,Herskind20}).
The second generation of solutions mainly focuses on improving unlinkability,
anonymity, and confidentiality, often considering decentralized architectures (e.g.,~\cite{miers13zerocoin,vanSaberhagen13cryptonote,Fauzi19:quisquis,baudet22:zef,friolo25,Sarencheh25:parscoin,Androulaki13,Herrera-Joancomarti15,Meiklejohn13}). 

We classify this second generation of approaches into three main categories:
privacy-native protocols,
mixing protocols,
and off-ledger solutions. 
Changing the protocol of existing blockchains is hard, as it involves forks. 
Therefore, \emph{privacy-native protocols} 
(mostly) defines novel solutions with improved confidentiality, anonymity, and unlinkability features (e.g.,~\cite{vanSaberhagen13cryptonote,noether16:monero,BenSasson14:zerocash,tomescu22:utt}). 
Conversely, \emph{mixing protocols} propose supplementary services for permissionless blockchains aimed to collect a group of transactions, scramble senders and receivers, and publish them as a single group transaction (e.g.,~\cite{Maxwell13:coinjoin,Bissias14:xim,Bonneau14:mixcoin,Valenta15:blindcoin,ruffing17valueshuffle,Ziegeldorf18coinparty,qin23:blindhub,Wang23}). Mixing protocols break the link between sender and receiver, thus reducing the possibility of linking together addresses of the blockchain ({an analysis of protocols in \refAppMixing{}}).
\emph{Off-ledger solutions} are a specific approach related to blockchains (that serve as Layer~1) and the ability to build decentralized peer-to-peer Layer~2 systems on top of them. Layer~2 solutions process transactions off the main chain, reducing the need to record every transaction on the shared ledger (e.g.,~\cite{qin23:blindhub,Wang23,baudet22:zef,seres19:mixeth}). 
Focusing on on-ledger privacy, we omit the analysis of Layer~2 privacy solutions.

\paragraph{Third Generation}
While protecting the privacy and anonymity of payers and payees remains essential, the third generation 
of solutions introduces mechanisms that balance these guarantees with \emph{auditability} and regulatory compliance 
(e.g.,~\cite{Cecchetti17,Garman17,Narula18,li21:traceable-monero,wust19:prcash,Zhang19,Androulaki20,chen20:pgc,Kiayias22:peredi,wust22:platypus,beer24:payoff,Papadoulis24:aqqua,Yu24,xue23}).
These approaches are motivated by concerns around AML and CFT.
Auditability refers to the system's ability to track, verify, or selectively disclose information about transactions, without fully compromising user privacy. This enables the enforcement of specific policies that apply to users, coins, or transactions behaviors over time. 
To this end, modern privacy-preserving systems integrate mechanisms such as revocable anonymity, trapdoors, and ZKPs for correctness and compliance.

\subsubsection{System Architecture}
We examine the architecture of privacy-preserving payment systems and protocols.

\paragraph{Centralized Systems}
The first generation of privacy preserving solutions typically rely on \emph{centralized architectures} involving up to four trusted authorities, each responsible for distinct functions. They include:
(i) a central bank, which issues digital coins and maintains monetary control;
(ii) a tracing authority, responsible for auditing and tracing under specific conditions;
(iii) a registration authority, in charge of user on-boarding and identity verification; 
and, in some cases, (iv) a technical support entity, which collaborates with the tracing authority to enable conditional de-anonymization. 
Most early proposals assume a single central bank (e.g.,~\cite{chaum83,dold19:taler,Cecchetti17,Zhang19,beer24:payoff}). 
For privacy, special cryptographic primitives are used, such as blind signatures~\cite{chaum83}, which allow the bank to issue valid coins without learning user identities.
To support auditability and regulatory compliance, systems often include a separate tracing authority that 
can decrypt transactions or enforce specific policies (e.g., check or bypass an anonymity budget).
This entity operates independently from the central bank, thereby preserving some separation of duties
(e.g.,~\cite{Sander99,wust22:platypus}). 
Yu et al.~\cite{Yu24} introduce a technical support authority that must cooperate with the tracing authority to extract sender identities. 
In the context of CBDCs, system architectures usually include a registration authority, responsible for enrolling user registration and managing identity records (e.g.,~\cite{Zhang19})---an essential requirement considering regulatory frameworks (e.g., AML and CFT).

\paragraph{Decentralized Systems}
Blockchains introduced a fundamental shift in system design by leveraging decentralization, 
which 
serves two main purposes. 
First, it enhances resilience against malicious actors or censorship by distributing trust across multiple entities.
Second, it enables intermediary-free value exchange, using commitments and digital signatures to ensure integrity, authenticity, and non-repudiation.
Many privacy-native solutions adopt decentralized architectures in line with these principles 
(e.g.,~\cite{noether16:monero,vanSaberhagen13cryptonote,poelstra16:mimblewimble,Androulaki20,Narula18,Ma21:dsc,xue23}).
Furthermore, the programmability of blockchains has enabled privacy-preserving overlays (e.g.,~\cite{miers13zerocoin,BenSasson14:zerocash})
and smart contracts\footnote{Although logically centralized, smart contracts are executed on each node; therefore, we consider them as a decentralized solution.} implementing private tokens ``pegged'' to the coin of the primary chain (e.g.,~\cite{bunz20:zether,baudet22:zef,Sarencheh25:parscoin}).

\subsubsection{Protocol Integration and Compatibility with Existing Blockchains}

Second-generation privacy-preserving payment systems vary in their integration with existing (blockchain) infrastructures: some remain compatible, others introduce non-compatible extensions, and some adopt entirely new designs.
 
\paragraph{Compatible Solutions}
Mixing protocols propose supplementary services for existing blockchains (mostly for Bitcoin), and are specifically designed to be fully compatible (e.g.,~\cite{ruffing16:coinshuffleplusplus,Valenta15:blindcoin,qin23:blindhub,ruffing17valueshuffle}).  
Another category of compatible solutions leverages blockchain programmability to introduce 
privacy-preserving
overlays (e.g.,~\cite{bunz20:zether,baudet22:zef}).
For instance, Zether~\cite{bunz20:zether} uses a smart contract on Ethereum to store and manage private accounts;  Zef~\cite{baudet22:zef} introduces opaque coins in the form of off-chain certificates exchanged between Ethereum user accounts.

\paragraph{Non-compatible Extensions}
Most privacy-enhancing proposals are non-compatible with the blockchain protocol they extend.
For example, Zerocoin~\cite{miers13zerocoin} is an overlay over Bitcoin that allows user to convert bitcoins in {zerocoins}. A zerocoin can be proved (in ZK) to originate from a valid and unspent bitcoin, without explicitly revealing such link. 
Zerocoin does not provide full anonymity: it obscures coin traceability, but it does not hide the number of transactions, recipient addresses, and balances.
Building on Zerocoin, Zerocash~\cite{BenSasson14:zerocash} enhances privacy by concealing both sender and recipient addresses, supporting arbitrary transaction values, and enabling change handling. 
Both Zerocoin and Zerocash require new transaction types and payment semantics, and break compatibility with Bitcoin.
\tlong{While, in theory, they could be integrated if a super-majority of miners and nodes adopt the new protocol, the authors do not expect such integration and instead recommend deploying Zerocash as a separate system. }Garman et al.~\cite{Garman17} extends Zerocash with no backward compatibility, by introducing policy enforcement and tracing of users and tainted coins---features that further diverge from Bitcoin's existing design.
Yu et al.~\cite{Yu24} propose an overlay on Bitcoin for anonymous yet regulated transactions using a coin backed by bitcoins. 
Their approach relies on three centralized entities \tlong{(i.e., bank, tracing authority, and technical support) }that cooperate to enable batch linkability and payer tracing while preserving privacy. 
These changes are not backward-compatible with Bitcoin.

\paragraph{Novel Approaches}
Most approaches present greenfield solutions for privacy-preserving payments or asset exchanges (e.g.,~\cite{vanSaberhagen13cryptonote,poelstra16:mimblewimble,noether16:monero,Narula18,Fauzi19:quisquis,wust19:prcash,Zhang19,Androulaki20,chen20:pgc}), often extending open-source solutions (e.g.,~\cite{li21:traceable-monero,Papadoulis24:aqqua}). 
Both centralized (e.g.,~\cite{Cecchetti17,Zhang19}) and decentralized architectures (e.g.,~\cite{Narula18,Androulaki20,lin20,Ma21:dsc,Goodell23,elfadul24}) exist, with a sensible preference for the latter.
Early approaches like CryptoNote~\cite{vanSaberhagen13cryptonote}, MimbleWimble~\cite{poelstra16:mimblewimble}, and Monero~\cite{noether16:monero} address Bitcoin's privacy limitations by still proposing fully decentralized alternatives.
CryptoNote~\cite{vanSaberhagen13cryptonote} enhances sender and receiver anonymity through one-time ring signatures and NIZK proofs, inspiring solutions like ByteCoin, DigitalNote, and Aeon~\cite{Amarasinghe19}.
MimbleWimble~\cite{poelstra16:mimblewimble} improves privacy and scalability by aggregating confidential transactions.\tlong{ However, this approach does not guarantee  unlinkability among participants who undertake the aggregation.}
Grin\footnote{\url{https://grin.mw/}} and Beam\footnote{\url{https://www.beam.mw/}} build on MimbleWimble ideas.
Monero~\cite{noether16:monero} strengthens privacy with sender, receiver, and network anonymity as well as confidential transactions.
\tlong{Despite its strong anonymity measures, research has identified vulnerabilities in Monero's mixing selection strategy that could lead to transaction de-anonymization (e.g.,~\cite{moser18,Kappos18}).}

Other approaches propose partially decentralized solutions (e.g.,~\cite{wust19:prcash}), combine concepts from centralized mixers (e.g.,~\cite{Fauzi19:quisquis}), or consider hierarchical bank-intermediated settings (e.g.,~\cite{Cecchetti17,Narula18}).
W\"ust et al.~\cite{wust19:prcash} propose PRCash to strike a trade-off between fast payments, user privacy, and regulatory oversight: it includes two centralized entities for issuance and oversight and a permissioned blockchain, with MimbleWimble-like transactions (i.e., commitment-based; see Section~\ref{sec:ppf:vr:su}).
Quisquis~\cite{Fauzi19:quisquis} builds on the concept of mixing to \emph{redistribute wealth}: each transaction updates a bunch of other accounts on chain, thus preventing addresses from appearing multiple times on the ledger, reducing the risk of de-anonymization and linkability. 
\cryptodef{Quisquis ensures integrity through ZKPs while maintaining privacy using commitments and updatable public keys.}

\subsubsection{Trust Assumptions}
\label{sec:ppf:trust}
We now focus on the degree of trust required
to facilitate secure transactions.
Most solutions for digital payments do not rely on trusted parties for confidentiality (e.g.,~\cite{Maxwell13:coinjoin,vanSaberhagen13cryptonote,noether16:monero,Narula18,Fauzi19:quisquis,li21:traceable-monero,Ma21:dsc,Guan22:blockmaze,tomescu22:utt,baudet22:zef,Nosouhi23:ucoin,Sarencheh25:parscoin,Wang23}), even when they propose a centralized system (e.g.,~\cite{chaum83,chaum21,wust22:platypus}).

The solutions that introduce trust assumptions do it for different purposes. 
The approaches using zk-SNARK as ZKP require a \emph{trusted setup} to initialize and correctly share some cryptographic parameters (e.g.,~\cite{miers13zerocoin,Garman17,BenSasson14:zerocash,xue23}).
A few works require a \emph{trusted central bank} to issue coins that can be later exchanged anonymously (e.g.,~\cite{Goodell23,friolo25}). 
Elfadul et al.~\cite{elfadul24} assume the presence of a regulatory supervision that grants permissions to users to perform transaction; this centralized authority could asymmetrically revoke such authorization. 
Contributions that explore special forms of auditability related to revocable anonymity (e.g.~\cite{wust19:prcash,lin20,xue23,Yu24}), trapdoor (e.g.,~\cite{Zhang19,Androulaki20,chen20:pgc,elfadul24}), or user tracing (e.g.,~\cite{Garman17}) require trust in the tracing authority that uncovers transaction details only when necessary.
PEReDi~\cite{Kiayias22:peredi} 
reduces trust requirements by relying on \emph{quorum} of centralized entities that must agree to perform specific operations (such as decrypting transactions).

\subsubsection{Interaction Models}
Privacy-preserving digital payment systems differ in whether transactions require direct interaction between payer and payee. 
\emph{Interactive payment protocols} need {active cooperation}  during the transaction process. This is common in account-based models where recipients help in constructing or approving the updated account state, via ZKPs or digital signatures.
For example, PEReDi~\cite{Kiayias22:peredi} and Platypus~\cite{wust22:platypus} enforce synchronized interaction: both parties participate in updating the ledger with proofs and commitments, ensuring mutual accountability and preventing attacks like dusting or replay. 
\tlong{These systems may offer enhanced control and auditability but at the cost of real-time coordination and increased communication overhead. }Similarly, Chaumian token systems often require immediate token claiming to prevent double spending (e.g.,~\cite{dold19:taler}). 
In contrast, \emph{non-interactive payment protocols} let the payer unilaterally create and broadcast transactions to the network.  
Many token-based systems, especially unspent token models, follow this approach (e.g.,~\cite{BenSasson14:zerocash,vanSaberhagen13cryptonote,noether16:monero,poelstra16:mimblewimble}).
This improves user convenience and enables asynchronous payments, although receivers may need  
periodic ledger scanning or local computation.\footnote{Off-chain communication can alleviate the burden of scanning the ledger, but current protocols lack this feature.} 
Some account-based protocols approximate non-interactive behavior via delayed claiming.
For instance, Sarencheh et al.~\cite{Sarencheh25:parscoin} propose a two stage and non-interactive variant of PEReDi:
first, the sender interacts with issuers to update his account (transfer stage); then, the receiver claims funds using enclosed ZKP and credits his account (claim stage).
Overall, interactive designs enable rich policy enforcement and conditional privacy,
whereas non-interactive models better suit high-latency or  asynchronous networks.

\subsection{Value Representation}
\label{sec:ppf:value}

The \emph{value representation} dimension examines how balances and transfers are modeled, focusing in particular on their privacy implications.

\subsubsection{Ledger Data Models} 
Digital payment systems and decentralized ledgers follow three main approaches for representing and tracking ownership of digital assets: accounts, tokens, and tabular ledger.

\paragraph{Account-based}
Account-based ledgers associate a balance with each account address, which is debited or credited by transactions. 
This simplifies state management and enables simple definitions of policies based on cumulative balances. Nonetheless, it requires safeguards against replay and double-spending attacks, often implemented by storing a sequence number (or \emph{nonce}) in each account and updating it with every transaction. 
Since the balance of an account is explicitly represented, achieving (pseudo-)anonymity and unlinkability is challenging, especially in systems where users cannot freely generate multiple addresses (e.g.,~\cite{Guan22:blockmaze}).
Privacy-preserving account-based solutions replace plain-text balances with commitments, achieving \emph{confidentiality} of balances and transactions (e.g.,~\cite{Fauzi19:quisquis,bunz20:zether,baudet22:zef,Guan22:blockmaze,Sarencheh25:parscoin})\footnote{While in the first generation of systems accounts are kept centralized and private, second and third generation systems are often public, enabling inspection and verification.}. 
Liu et al.~\cite{liu22} propose an account-based CBDC with sender and receiver anonymity but with no unlinkability: The system follows a two-tier model where the central bank can identify users via encrypted identifiers, while commercial banks cannot\tlong{:
users use blind signatures by the central bank for account generation and withdrawals and prove validity of their operations using ZKPs}.
To guarantee 
anonymity while ensuring \emph{unlinkability}, most of the approaches change users' account address at each transaction and use an authenticator to prove correctness. The authenticator can be implemented, e.g., using digital signatures by trusted or centralized authorities (e.g.,~\cite{Kiayias22:peredi}) or ZKPs (e.g.,~\cite{wust22:platypus,Fauzi19:quisquis,bunz20:zether,Papadoulis24:aqqua}). 
Platypus~\cite{wust22:platypus} rely on ZKPs; 
conversely, in PEReDi~\cite{Kiayias22:peredi}, account states must be (blindly) signed by a quorum of maintainers, and since each transaction reveals prior account snapshots to construct new authenticators, randomized signatures are used to prevent linkability.
Quisquis~\cite{Fauzi19:quisquis} and
AQQUA~\cite{Papadoulis24:aqqua} use updatable public keys
to update accounts in a decentralized manner.
The account-based model lends itself well to \emph{auditability} implemented in third generation systems, as it allows simple definition of policies on holding limits (e.g.,~\cite{gross2021,wust22:platypus,Papadoulis24:aqqua}) or spending and receiving limits (e.g.,~\cite{Papadoulis24:aqqua}).
Gross et al.~\cite{gross2021} includes, in the account state, an accumulator of 
all spending in the current epoch, enabling policies on this information.
AQQUA~\cite{Papadoulis24:aqqua} includes commitments to three different counters (i.e., balance, total amounts sent and received). 
Conversely, Platypus~\cite{wust22:platypus} leaves the specification of the additional information to be application dependent.

\paragraph{Token-based}
Token-based ledgers do not maintain explicit balances per address, but instead track individual tokens. The user's balance can be derived from the sum of all owned tokens. This model enables parallel validation of user transactions and stronger privacy guarantees (as balances are not explicitly materialized) but introduces management complexity, particularly when users distribute their coins across multiple addresses for added privacy.
There are two main types of token structures: {Chaumian tokens} and unspent tokens.

Chaumian tokens~\cite{chaum83} use blind signatures and were first deployed in centralized, bank-issued electronic cash schemes (i.e., first generation systems).
When issuing a token, the bank blindly signs a random value provided by the payer, allowing the payer to later unblind it for spending (ensuring \emph{sender anonymity}). 
These tokens rely on a trusted authority to issue and verify their validity; to this end, the authority must hold an ever increasing list of spent tokens that might introduce scalability challenges.
Some implementations introduce expiration to limit circulation (e.g.,~\cite{dold19:taler,lovejoy23:hamilton}).
Chaumian-inspired decentralized privacy-preserving systems of the second generation (e.g., Monero~\cite{noether16:monero}, Zerocoin~\cite{miers13zerocoin}, and Zerocash~\cite{BenSasson14:zerocash}) eliminate the need of a central issuer using a combination of different cryptographic techniques. 
For example, Zerocash supports transparent (like Bitcoin) and shielded transactions, with the latter hiding sender, recipient, and amount.
Shielded payments encrypt these values with one-time keys and use zk-SNARK proofs to validate inputs-outputs equality, concealing all on-chain details.

Unspent tokens, introduced in Bitcoin~\cite{nakamoto08bitcoin}, represent transaction outputs that can be used as inputs for future transactions, which in turn create new unspent tokens. As such, they are inherently traceable unless PETs are used~\cite{Baldimtsi24:sok}. 
Unspent tokens are issued in a decentralized manner, usually as result of the mining process.
Their core attributes include an identifier\tlong{ (e.g., transaction hash and output index)}, value (or its commitment), and ownership information.
Their recording in a public, transparent ledger prevents double-spending but complicates privacy.
Privacy-focused crypto-asset systems employ stealth addresses, ring signatures, confidential transactions, and ZKPs to \emph{obscure} transaction details and \emph{unlink} payments (e.g.,~\cite{noether16:monero,poelstra16:mimblewimble,miers13zerocoin,BenSasson14:zerocash})---these properties are discussed in more detail later.
Tokens can also include auxiliary data to support the evaluation of specific \emph{policies} (e.g.,~\cite{xue23,Yu24}) for third generation systems. 
For example, Xue et al.~\cite{xue23} include data to prove in ZK that the total 
value transferred and the transaction frequency in a time period is limited.
Section~\ref{sec:ppf:audit} further discusses the auditability of token-based systems.

\paragraph{Tabular Ledger}
{Narula et al.~\cite{Narula18} propose zkLedger, a third approach tailored to a specific use case: a bank-intermediated CBDC.
It uses a tabular ledger where rows represent transactions and columns represent banks.
Each cell contains a Pedersen commitment to hide values,
while enabling cryptographic verification. 
Transactions include ZKPs to 
prove balance, asset ownership, and consistency with audit records.
This structure supports rich privacy-preserving \emph{auditing}: this third generation system includes an auditor who can issue a rich set of auditing queries---using primitives like sums, moving averages, variance, standard deviation, ratios---and receive answers that are provably consistent with the ledger.
As a downside, the tabular structure scales poorly with the number of banks and transactions, since the ledger grows monotonically.
MiniLedger~\cite{Chatzigiannis21} adopts a similar tabular abstraction but addresses scalability by replacing commitments with (twisted ElGamal) encryptions and introducing cryptographically verifiable ledger pruning. 
MiniLedger allows old transaction rows to be compacted into digests without losing auditability: banks can later selectively disclose individual transactions, aggregates, or balances together with ZKPs, even after pruning. This makes MiniLedger better suited for long-running inter-bank settlement scenarios.

\subsubsection{Ledger State Update} 
\label{sec:ppf:vr:su}
Updating the ledger may reveal sensitive information when it exposes which state elements are created, updated, or consumed 
to non-involved parties (e.g., globally on public ledgers or to intermediaries), enabling linkability through update patterns (even if values are hidden).
Privacy-preserving systems rely on commitments, cryptographic accumulators (e.g., Merkle trees), or tags such as nullifiers, to validate state transitions and prevent double spending while minimizing information leakage.

\paragraph{Account-based}
A generic state update transaction for account-based ledger includes the sender and receiver addresses, the exchanged amount, and the sender digital signature. 
For privacy-enhanced account-based, a na\"{i}ve state update would reveal the updated accounts.
Therefore, privacy-preserving state updates in account-based ledger usually adopt three strategies  (e.g.,~\cite{Kiayias22:peredi,wust22:platypus,Fauzi19:quisquis,bunz20:zether,Papadoulis24:aqqua}): to obtain confidentiality, they update a commitment to the new account state; to obtain unlinkability, they always update the public key associated with the (recipient) account; and to obtain anonymity, they update a group of accounts at the same time.

\paragraph{Token-based}
Token-based ledgers store the list of valid (unspent) tokens or burnt (spent) tokens. 
A spending operation must invalidate an existing unspent token, to prevent double-spending. 
With Chaumian tokens (e.g.,~\cite{chaum83,dold19:taler,lovejoy23:hamilton}), payment systems include a withdrawal operation\tlong{ (or mint transaction)}, which creates a new valid token that the payer can spend later in time, and a spend operation\tlong{ (or pour transaction)}. The state of the system is represented by the list of spent tokens, which is updated at spending time; unspent tokens are usually not explicitly stored (if not by payers).
Privacy-preserving systems based on Chaumian tokens do not usually guarantee transaction confidentiality (e.g.,~\cite{chaum83,hayes90,Sander99,dold19:taler,Goodell23,friolo25}), therefore they still maintain the list of spent tokens.
Unlike other approaches, Sander and Ta-Shma~\cite{Sander99} represent valid tokens in a Merkle tree. The bank does not blind sign tokens; instead, valid tokens come with a hash chain proving membership in the Merkle tree. The Merkle root is publicly available to ensure transparency and auditability. Withdrawal coins can be invalidated if needed, addressing certain attack scenarios (e.g., blackmail).
\tlong{To hide amounts, Monero wraps each amount in a Pedersen commitment, and proves via a Bulletproof range ZKP that each value lies in a valid range (no negative or overflow). The additive properties of Pedersen commitments allows proving input-output equality without revealing any value. }Friolo et al.~\cite{friolo25} employ a ZKP-based protocol to achieve \emph{unlinkability}: Transactions are represented as commitments, and when spent, their opening is revealed along with a ZKP demonstrating that it is linked to a valid, unspent token.
In contrast, UTT~\cite{tomescu22:utt} ensures \emph{transaction confidentiality}, and it recurs to \emph{nullifiers} to track spent coins. Nullifiers enable the bank to detect double-spending without knowing transaction specifics. 
In~\cite{tomescu22:utt}, a nullifier is a pseudo-random value computed from the coin's serial number 
(and a secret).
The system checks their validity
through ZKPs and, if valid, stores them in a list for future checks.

With unspent tokens, the payment operation simultaneously invalidates previous tokens and generate new ones. The state of the system is represented by the list of unspent tokens\footnote{Since blockchains store transactions in an append-only log, they trace the derivation history of each unspent token.}.
Unspent tokens cannot be easily marked as spent without compromising \emph{sender anonymity}.
Most of the approaches, including CryptoNote~\cite{vanSaberhagen13cryptonote}, Zerocoin~\cite{miers13zerocoin}, Zerocash~\cite{BenSasson14:zerocash}, and their extensions (e.g.,~\cite{Garman17,Androulaki14,xue23}), let the payer spend a token by using a serial number and a ZKP that indicates that the number is a valid opening to some commitment stored in an unspent token. 
Monero~\cite{noether16:monero} (and its extension Traceable Monero~\cite{li21:traceable-monero}) uses one-time signatures for unspent tokens, which can be used only once to spend the token. At spending time, the ledger registers a salted hash of the public key. 
MimbleWimble (and PRCash that reuses its transaction format~\cite{wust19:prcash}) 
uses a binding and hiding commitment scheme~\cite{Pedersen92}, together with Bulletproof range proofs~\cite{bunz18:bulletproof}, to define the recipient.
By using commitments as outputs, coins act as having their own private key, the knowledge of which, along with the knowledge of their value, enable spending the token. 
This guarantees \emph{anonymity} and \emph{unlinkability}.

\subsubsection{Handling Change} 
Account-based ledgers use transactions to transfer exact value between payer and payee, therefore these kind of systems do not need to support change.
Instead, token-based ledgers propose different approaches for handling change. 
Chaumian tokens do not easily support change or partial spending (e.g.,~\cite{dold19:taler}).
Tomescu et al.~\cite{tomescu22:utt} allow multiple inputs and outputs for a single transaction, but requires the payer to prove \emph{value preservation} in ZK. The proof consists in two parts: the first proves that the sum of inputs equals the sum of outputs; the second is a range proof on the output value (needed because ZKPs works with modulo arithmetic and the equality proof is not enough to guarantee value preservation).
System based on unspent tokens inherit the change implementation strategy by Bitcoin, where the change is another unspent token owned by a (new) address managed by the payer (e.g.,~\cite{miers13zerocoin,Androulaki20,elfadul24}) or a commitment controlled by the payer (e.g.,~\cite{noether16:monero,wust19:prcash}).

\subsubsection{Support for Different Denominations} 
Chaum~\cite{chaum83} considered a simple system where tokens have implicit value. 
A few extensions consider the possibility to emit tokens with different denominations (e.g.,~\cite{dold19:taler,tomescu22:utt}). For example, Dold~\cite{dold19:taler} in GNU Taler uses multiple denomination key pairs for blind-signing coins of different monetary values. Conversely, Tomescu et al.~\cite{tomescu22:utt} explicitly represent the token value by concealing it within a commitment.
As regards unspent token-based ledgers, Zerocoin does not support arbitrary denominations~\cite{miers13zerocoin}. This design simplifies the ZKP used for anonymity but introduces usability limitations, as users must split or combine coins when making payments.  
Zerocash~\cite{BenSasson14:zerocash}
supports arbitrary denominations by using homomorphic commitments, allowing users to transact any amount while preserving privacy.
The remaining approaches allow tokens of different denominations, as each unspent token internally holds its value (or a commitment to it).
Among them, CryptoNote~\cite{vanSaberhagen13cryptonote} does not hide transaction amounts, making it vulnerable to statistical analysis---e.g., identifying change as the smaller output.
Using multiple change addresses can obscure amounts but leads to proliferation of \emph{dust  transactions}~\cite{noether16:monero}. 
Hiding values with commitments prevents such analysis.

\subsubsection{Token Life-cycle}
\label{sec:ppf:tlc}

\paragraph{Double Spending Protection}
Two approaches protect from double spending: deposit-time checks
and publicly verifiable ledgers.
Centralized systems with Chaumian tokens do not usually rely on a transparent ledger;
instead, they check token validity 
at deposit time (e.g.,~\cite{chaum83,dold19:taler,tomescu22:utt,Goodell23}). 
The public visibility of the ledger in blockchains (e.g.,~\cite{nakamoto08bitcoin}) allows payee to autonomously check token validity. This is also possible in privacy preserving systems, which usually require the payer to provide a ZKP of token validity (e.g.,~\cite{miers13zerocoin,poelstra16:mimblewimble,noether16:monero,li21:traceable-monero,xue23,Yu24}).

\paragraph{Transferable Tokens}

Few works (e.g.,\cite{hayes90,Goodell23}) propose mechanisms for transferring Chaumian tokens. Hayes~\cite{hayes90} proposed chaining transactions, where each owner signs over the previous link; long chains can be refreshed by the bank, though the scheme risks double spending, mitigated by token lifetimes.
Goodell et al.~\cite{Goodell23} instead attach a proof of provenance---a list of commitments with the issuer's signature---allowing validity checks without external interaction.

\paragraph{Immediate Claiming}
With Chaumian tokens, transactions occur off-ledger, so payees must deposit tokens promptly to prevent double spending~\cite{chaum83}. Misbehavior can be deterred by one-time signatures that reveal the sender's identity~\cite{hayes90,Camenisch06}. Sander and Ta-Shma~\cite{Sander99} instead use Merkle trees: coin validity is proven via inclusion proofs to public roots, enabling offline asynchronous payments without real-time bank interaction.

\subsection{Privacy Features}
\label{sec:ppf:privacy}
Privacy in digital payment systems is often defined through several sub-properties, including participant anonymity, value confidentiality, plausible deniability, and unlinkability (e.g.,~\cite{Almashaqbeh22,Amarasinghe19,Bernal19,genkin18}). 
The privacy mechanisms discussed in this section are largely shared across generations; differences arise primarily in how these mechanisms are combined and integrated with system-level objectives, with 
techniques such as encryption, ZKPs, and credentials\footnote{By credentials we refer to cryptographic constructions---typically based on digital signatures combined with ZKPs---that bind attributes or rights to a user and enable selective disclosure without revealing unnecessary information.} 
enabling selective disclosure and auditability.

\subsubsection{Sender Anonymity}
\label{sec:ppf:sa}

\emph{Asymmetric anonymity} indicates that 
only one of the participants' identities (sender or receiver) is protected
(e.g.,~\cite{chaum83,Sander99,miers13zerocoin,dold19:taler,Zhang19,bis23:tourbillon,Goodell23,Yu24,Androulaki14}).
\emph{Sender anonymity} focuses only on hiding the payer's identity, usually using the following techniques:
blind signatures (e.g.,~\cite{chaum83,dold19:taler,Goodell23,liu22}) and their variants---one-time (e.g.,~\cite{hayes90}), threshold (e.g.,~\cite{Sarencheh25:parscoin,Kiayias22:peredi,Androulaki20}), and randomizable (e.g.,~\cite{tomescu22:utt,wust19:prcash})---, ring signatures (e.g.,~\cite{vanSaberhagen13cryptonote,noether16:monero,elfadul24}), group signatures (e.g.,~\cite{Zhang19}), and commitments with ZKPs (e.g.,~\cite{miers13zerocoin,BenSasson14:zerocash}).

Chaum~\cite{chaum83} proposes \emph{blind signatures} to unlink the sender request of token issuance and its later spending.
To reveal the sender on double spending, Hayes~\cite{hayes90} introduces \emph{one-time blind signatures}. However, since each transaction requires a new signing key, storage and computation management become cumbersome; moreover, one-time signatures often require large signature size and expensive verification, making them impractical for high-throughput systems. 
UTT~\cite{tomescu22:utt} implements blind signatures using 
\emph{randomizable signatures}~\cite{pointcheval16:randsig},
which allow signatures to be 
re-randomized using fresh randomness without compromising their validity. 
\tlong{In particular, each coin consists of a commitment signed by the central bank, where the commitment binds the payer's public identifier, the coin value, and a random nonce (to ensure hiding). }All these schemes rely on a central issuer, which represents a potential single point of failure. This can be mitigated, e.g., by using \emph{threshold blind signatures}, where a quorum of authorities must cooperate to create a valid signature (e.g.,~\cite{Androulaki20,Kiayias22:peredi,Sarencheh22:peredi:techrep}).

Other works reuse the idea of mixing through \emph{ring signatures}, which conceals the sender's public key within a group of other public keys---the \emph{anonymity set} (e.g.,~\cite{vanSaberhagen13cryptonote,noether16:monero,li21:traceable-monero,elfadul24}). 
For example, CryptoNote~\cite{vanSaberhagen13cryptonote} employs one-time ring signatures: each transaction uses a fresh key and produces a key image, enabling detection of double spends. 
Different privacy-preserving crypto-asset systems use ring signatures for obtaining sender anonymity, including Monero\footnote{M\"oser et al.~\cite{moser18} showed weaknesses in Monero's key sampling: $66.09\%$ of transactions do not include any other public key, sampling is not really random, and real inputs can be inferred with $80\%$ accuracy.}~\cite{noether16:monero}, Traceable Monero~\cite{li21:traceable-monero}, and PCRAP~\cite{elfadul24}. 
{However, studies show that Monero's anonymity can be compromised through transaction analysis (e.g.,~\cite{moser18,Kappos18,kumar17}).}

To revoke sender anonymity if needed, Zhang et al.~\cite{Zhang19} use \emph{group signatures} where a trusted centralized entity sets up the anonymity set and can, optionally, uncover the senders' identity for inspection.

To be valid, unspent tokens require proving their ownership. To implement sender anonymity, existing approaches rely on two key techniques. 
First, the unspent token includes a commitment to its serial number; the serial number is revealed only to mark the token as spent (see Section~\ref{sec:ppf:tlc}). 
Second, the spending operation includes a ZKP that demonstrates knowledge of the opening to a commitment of an unspent coin. 
Sander and Ta-Shma~\cite{Sander99} use multiple Merkle trees to track valid Chaumian tokens. To prove validity, the sender provides a ZKP attesting knowledge of a valid coin in the tree and of the coin's serial number and a hash chain leading to one of Merkle tree roots.
Zerocoin~\cite{miers13zerocoin},  
and its extensions (e.g.,~\cite{BenSasson14:zerocash,Garman17}), 
adopts ZKPs as well: the payer publishes a transaction that includes the recipient's address, an empty origin address, a ZKP, and the commitment opening. 
The ZKP prevents linking the sender to any specific commitment; the commitment itself hides the opening using a random value that is never revealed, preventing any association between the two.

Privacy-preserving account-based models mostly rely on ZKPs to obtain sender anonymity (e.g.,~\cite{gross2021,Kiayias22:peredi,wust22:platypus,beer24:payoff}).
Specifically, a commitment to the account state is used, with updates including ZKPs to prove correctness.
For example, Gross et al.~\cite{gross2021} use the ideas of Zerocash by committing on the accounts state and by invalidating them on update. The authors suggest using Merkle trees to prove that a transaction proposal refers to an existing commitment in the ledger without pointing to (and thus revealing) it.\footnote{For a broader discussion of set accumulators---including Merkle trees as a key example---and their role in enhancing anonymity, see~\cite{Loporchio23:survey-accum}.} 
Platypus~\cite{wust22:platypus} uses a model where payer and payee collaborate to exchange with the central bank commitments to their updated accounts, the transaction (amount and  serial number), and two ZKPs proving the update's validity and its link to a prior certified state. 
If valid
the bank updates a public ledger.
Conversely, PEReDi~\cite{Kiayias22:peredi} achieves sender anonymity through anonymous channels, an encrypted ledger, and a threshold signature scheme. 
Sender and receiver submit partial (specular) transactions that update only their respective account states, 
encrypted using threshold encryption.
Solidus~\cite{Cecchetti17} is a hierarchical system where banks act as proxies, managing on-chain transactions on behalf of users
and ensuring transaction-graph confidentiality.\footnote{While banks can identify their respective clients, other entities only learn the identities of the banks.}

\subsubsection{Receiver Anonymity}
The key techniques for receiver anonymity are one-time destination keys or stealth address (e.g.,~\cite{vanSaberhagen13cryptonote,noether16:monero,elfadul24,Zhang24}), encrypted address (e.g.,~\cite{BenSasson14:zerocash,tomescu22:utt}), commitments as output (e.g.,~\cite{poelstra16:mimblewimble,wust19:prcash}), and updatable public key (e.g.,~\cite{Fauzi19:quisquis,Papadoulis24:aqqua}).
Obtaining receiver anonymity with Chaumian tokens is challenging, as the payee must interact with the bank to deposit the received tokens. 
Hayes~\cite{hayes90} suggests using \emph{one-time aliases} to obtain privacy, even though the idea is not detailed for practical usages.
A similar idea can be found in CryptoNote~\cite{vanSaberhagen13cryptonote}, where the sender generates a \emph{one-time destination key} by combining a random number with the receiver's public key---essentially performing a Diffie-Hellman exchange to derive a shared secret.\footnote{Because this process requires two elliptic curve keys, a CryptoNote address is twice the size of a standard elliptic curve public key.}
The sender posts the transaction to a one-time public key with a commitment to a random value. The receiver scans all transactions with his private key, reconstructs the destination key, and accepts the payment if it matches. Sender and receiver never interact directly, but each receiver must scan the entire ledger to find his payments.
This approach has been later integrated in other systems, including Monero~\cite{noether16:monero}, which calls the one-time receiver address as \emph{stealth address}, Traceable Monero~\cite{li21:traceable-monero}, and PCRAP~\cite{elfadul24} (see Section~\ref{sec:prot:otpk}). 

In Zerocash~\cite{BenSasson14:zerocash} (unspent token-based ledger), 
the sender includes an \emph{encrypted version of the recipient's address} in the transaction (using asymmetric encryption). 
Therefore, periodically, each user has to 
scan the ledger, decrypt transactions using his private key, and identify those intended for him. 
\tlong{This prevents external observers from linking payments to specific recipients while allowing recipients to identify incoming transactions without revealing their addresses publicly. }UTT~\cite{tomescu22:utt} (Chaumian token-based system) ensures receiver anonymity using \emph{homomorphic commitments} and anonymous IBE scheme~\cite{boneh:ibe}. 
A coin is a commitment to a tuple containing the owner's identifier, a serial number issued by the central bank, and its value. To be valid, each coin comes with a (randomizable) signature by the central bank. 
To make a payment, the sender creates a transaction for the payee, 
which includes the receiver's identity commitment, the output value commitment, the encrypted coin details (with the recipient's key), and ZKPs.
The transaction is sent to the bank for validation; if valid, the bank blindly issues new coins for the recipient and records them on a public ledger\footnote{The homomorphic properties of the commitment scheme ensures that the bank does not learn identities or amounts, and allows to directly combine the value commitment, identity commitment, and serial number commitment in a single coin commitment.}. 
Recipients periodically scan the ledger to get their coins, to be randomized before adding to the wallet.

The unspent token-based ledger MimbleWimble~\cite{poelstra16:mimblewimble} (and PRCash~\cite{wust19:prcash}) performs transactions between \emph{commitments} rather than  addresses;
this guarantees both receiver anonymity and unlinkability.

Account-based approaches typically achieve receiver anonymity using the same technique as for sender anonymity (e.g.,~\cite{wust22:platypus, Kiayias22:peredi, Cecchetti17, bunz20:zether, lin20, gross2021, baudet22:zef, Guan22:blockmaze, beer24:payoff, Sarencheh25:parscoin, Papadoulis24:aqqua, Fauzi19:quisquis}). This is done by \emph{updating account state commitments} and employing ZKPs to verify correctness.
We mention Quisquis~\cite{Fauzi19:quisquis}, 
where an account is a pair of public key and a homomorphic commitment to its balance. A transaction updates both the public key and the commitment using 
\emph{updatable public key} (see Section~\ref{sec:prot:otpk}).
Senders select public keys---including the recipient's and decoys---to form the transaction input, updating only the sender's and recipient's balances while leaving others unchanged. 
Since each address appears at most twice on the ledger, de-anonymization is harder. However, this approach requires users to periodically scan the ledger to update their keys, thus introducing some transaction serialization.

\subsubsection{Transaction Value Confidentiality}
Two main approaches conceal transaction details:
commitments (e.g.,~\cite{BenSasson14:zerocash,Androulaki14,noether16:monero,poelstra16:mimblewimble,Garman17,Fauzi19:quisquis,wust19:prcash,gross2021,Guan22:blockmaze,wust22:platypus,beer24:payoff,elfadul24,tomescu22:utt,Androulaki20,Narula18})
and public key encryption (e.g.,~\cite{Kiayias22:peredi,Sarencheh25:parscoin,bunz20:zether,Ma21:dsc}).

\paragraph{Commitments}
Camenisch et al.~\cite{Camenisch06} use commitments with verifiable encryption to hide the commitment's opening. 
In 2013, Maxwell proposed Confidential Transactions\footnote{\url{https://elementsproject.org/features/confidential-transactions/investigation}}, which use \emph{Pedersen commitments} to hide transaction amounts while ensuring correctness via range proofs (e.g., Bulletproofs). 
Monero combines this technique with ring signatures obtaining RingCT~\cite{noether16:monero}.
Zerocash~\cite{BenSasson14:zerocash} uses {ZKPs} 
to achieve full transaction privacy: a cryptographic commitment is made to a new coin, including its value, owner, and serial number; then, a ZKP ensures transaction validity without exposing sensitive data.
Commitment-based schemes are the most widely adopted to obtain confidential transaction value, especially because their homomorphic versions allow to combine commitments without leaking sensitive data. 
To prevent users from creating negative or infinite amounts, range proofs prove that a committed value is greater than zero and within a valid range.

\paragraph{Public Key Encryption}
In PEReDi~\cite{Kiayias22:peredi}, senders and receivers update their accounts by submitting  \emph{encrypted transactions} to several maintainers who operate a distributed ledger. 
Threshold ElGamal encryption~\cite{elgamal85}  requires a quorum of maintainers to cooperate to decrypt transaction details.
Authorities can trace both transactions and identities. 
PARScoin~\cite{Sarencheh25:parscoin} extends this approach to introduce an offline payment protocol that does not require interaction between sender and receiver. 
In Solidus~\cite{Cecchetti17}, Publicly Verifiable Oblivious RAMs (PVORMs) enable banks to update the accounts of their clients without revealing exactly which accounts are being updated. PVORM provides a ZKP demonstrating that the updates are correct with respect to the transaction triggering them.

A few solutions implement privacy-preserving digital payments using smart contracts on public blockchains. To ensure confidentiality, transactions are encrypted with the sender's public key and supplemented with a NIZK proof for public verifiability (e.g.,~\cite{Ma21:dsc, bunz20:zether}).
For example, Zether~\cite{bunz20:zether} uses smart contracts to define privacy-preserving tokens that can be exchanged between public keys associated with a confidential balance. 
They use ElGamal encryption, which has homomorphic properties, to hide each account's balance.
Zether introduces the $\Sigma$-Bullets (an enhancement of the Bulletproofs range ZKP) to efficiently prove statements over the encrypted transfer value and the new sender balance. 
However, Zether currently allows only one anonymous transfer per epoch (to prevent double spending), and execution costs (namely, gas costs on Ethereum) are very high~\cite{Almashaqbeh22}. 
Similarly, DSC~\cite{Ma21:dsc} optimizes privacy within Ethereum-style smart contract environments, but does not support anonymity: it uses ZKPs and homomorphic encryption to hide balances and transferred amounts. User balances are stored on the ledger in encrypted form. 
\tlong{The homomorphic properties allows validators to update the balance in ciphertext without the need of decryption.}

\subsubsection{Network Anonymity}
\label{sec:ppf:net}
Network anonymity concerns the protection of sender and receiver IP addresses, which could otherwise reveal real-world identities.
Most of the existing approaches do not investigate (or even mention) the network anonymity feature. 
Few works mention it but consider this issue  as out of scope of their investigation (e.g.,~\cite{Cecchetti17,baudet22:zef,tomescu22:utt,wust22:platypus}). 
\tlong{They consider this feature as orthogonal hence that can be achieved by implementing the digital payment system on top of a private overlay network. }Monero~\cite{noether16:monero} and Traceable Monero~\cite{li21:traceable-monero} explicitly support network anonymity by using TOR (The Onion Router\footnote{\url{https://www.torproject.org/}}) and I2P (Invisible Internet Project\footnote{\url{https://geti2p.net/en/}}) to hide the real IP address~\cite{shi24:deanonym-monero}.
While \emph{TOR} routes traffic through multiple relays to obscure the user's IP address, \emph{I2P} is a network optimized for anonymous peer-to-peer communication, using fully decentralized routing and encryption to enable different anonymous services.
Moreover, Monero uses 
\emph{Dandelion++}~\cite{fanti18:dandelionplusplus}, an improvement of Dandelion~\cite{Venkatakrishnan17:dandelion}, which prevents attackers from linking transactions
to their source IP addresses. 
Dandelion++ uses randomization mechanisms for message forwarding, which happens in two phases: stem (anonymization) and fluff (broadcast). 
In the stem phase, a transaction is first relayed privately across a small, randomly chosen subset of nodes; each node probabilistically decides whether to continue forwarding the transaction along the stem or to switch to the second phase. 
In the fluff phase, the node broadcasts to the entire network using traditional flood propagation protocol.

\subsubsection{Unlinkability}
\label{ppf:aud:unlinkability}
Unlinkability ensures that transactions cannot be correlated or traced back to the same user (other definitions in~\cite{chaum83,Amarasinghe19,zhang23}).
It can refer to the sender, receiver, or both. 
Obtaining unlinkability in public ledgers is challenging as links between addresses in crypto-asset systems stems from different patterns (e.g.,~\cite{ron13:txgranalysis,chan17icitst,beres21:dapps,victor20:fc,zhong24:kdd,lu22:europlop,ermilov17icmla,kalodner20security});
a non-exhaustive list follows.

\begin{itemize}
    \item \emph{Common input ownership heuristic}: If multiple addresses appear as inputs in a single transaction, they are likely controlled by the same entity;

    \item \emph{Change address}: In a transaction with two outputs, one is the payment, the other one may be the change returned to the sender;

    \item \emph{Address reuse}: Using the same address for multiple transactions makes it easy to link payments to the same user\footnote{This works because address are pseudonymous but not private.};

    \item \emph{Temporal analysis}: If transactions occur in rapid succession or at predictable intervals, they may be linked to the same entity\tlong{, e.g., due to automated systems that process transactions in batches or periodically};
    
    \item \emph{Clustering analysis}: It deals with identifying wallet behavior such as round numbers, specific fee patterns, or common spending habits;

    \item \emph{Dust attack analysis}: Small amounts of crypto-assets are sent to many addresses to track their movements (similar to a phishing attack); if dust is later spent with other transactions, it links addresses together.

\end{itemize}
Existing countermeasures primarily aim to obscure input-output relations within transactions by using mixing protocols and preventing address reuse.
Mixing protocols disrupt common-ownership heuristics and help mitigate change-address attacks (an analysis of mixing protocols in ~\refAppMixing{}).
Address-changing mechanisms, such as generating a fresh sender or receiver address for each transaction, further reduce linkability.
In addition, using network-level anonymity tools helps conceal user identities. 
Finally, the simplest defense against dust attacks is to ignore dust transactions, thereby preventing adversaries from linking addresses.

\paragraph{Integrated Solutions}
Privacy-native solutions with public ledgers leverage a variety of strategies to improve unlinkability, focusing primarily on address-changing, transaction mixing, and use of ZKPs. Each of these techniques adds a layer of privacy, with trade-offs in terms of anonymity, computational complexity, and vulnerability to analysis.
One-time destination addresses, \emph{stealth addresses} (e.g.,~\cite{vanSaberhagen13cryptonote,noether16:monero}), or updatable public keys (e.g.,~\cite{Papadoulis24:aqqua,Fauzi19:quisquis}) help obtain recipient unlinkability. While this ensures the recipient's privacy, it does not hide transaction amounts.
With \emph{ring signatures} (e.g.,~\cite{noether16:monero,vanSaberhagen13cryptonote}),
the privacy protection is limited by the number of decoys in the ring, and sophisticated heuristics can sometimes reduce the anonymity set.
For the most robust privacy, \emph{ZKPs} offer a powerful solution (e.g.,~\cite{BenSasson14:zerocash,poelstra16:mimblewimble}). They allow transactions to prove their validity without revealing any details about the sender, receiver, or transaction amount. 
With this approach, transactions are recorded as cryptographic proofs (instead of sender-receiver addresses), ensuring that outputs cannot be linked to inputs.
Among ZK-based approaches, we mention the \emph{commitments and nullifiers} strategy (e.g.,~\cite{BenSasson14:zerocash,tomescu22:utt}):
a coin contains a commitment to its serial number, and spending the coin involves publishing a nullifier, i.e., a pseudo-random value derived from the coin serial number.
This ensures unlinkability to the original coin, while marking it as spent.
However, designing a system with commitments and nullifiers is more complex than 
ring signatures or stealth addresses, and is also more computationally demanding (see Section~\ref{sec:ppf:summary}).

\subsubsection{Plausible Deniability}
\emph{Plausible deniability} indicates whether users can deny intentional participation in a transaction, preventing others from proving their involvement (e.g.,~\cite{Almashaqbeh22,Ziegeldorf18coinparty}).
Existing works are roughly evenly divided between those supporting this property (e.g.,~\cite{noether16:monero,poelstra16:mimblewimble,Narula18,Kiayias22:peredi,tomescu22:utt,wust22:platypus,Goodell23,Papadoulis24:aqqua}) and those not (e.g.,~\cite{Maxwell13:coinjoin,vanSaberhagen13cryptonote,Garman17,bunz20:zether,gross2021,baudet22:zef,Guan22:blockmaze,Yu24}).
Basically, protocols ``on-demand'', such as mixing protocols, subsume an explicit user intention in using the additional service to obtain a privacy feature (such as unlinkability). Likewise, overlay systems implementing anonymous crypto-asset systems, such as Zerocoin~\cite{miers13zerocoin}, Zerocash~\cite{BenSasson14:zerocash}, EZC~\cite{Androulaki14}, Zether~\cite{bunz20:zether}, require explicit intention to convert a native asset to a privacy-preserving one. 
Conversely, systems that force anonymity or unlinkability as default, readily provide the property of plausible deniability. Among these systems, we mention Monero~\cite{noether16:monero}, MimbleWimble~\cite{poelstra16:mimblewimble}, UTT~\cite{tomescu22:utt}, PEReDi~\cite{Kiayias22:peredi}, or Platypus~\cite{wust22:platypus}, that do not provide the functionality of combining anonymous and public transactions.

\subsection{Auditability}
\label{sec:ppf:audit}

While privacy is a core requirement of digital payment systems, complete opacity is often incompatible with regulatory and operational constraints.
In regulated environments (e.g., CBDCs), some form of oversight is required to support AML/CFT controls, institutional accountability, and post-incident analysis (e.g., fraud, theft, protocol failures).
Auditability should not be conflated with full transparency or continuous surveillance. 
Prior work has highlighted a broad design space between cash-like anonymity and full transaction observability, in which selective oversight can be enabled without pervasive disclosure (e.g.,~\cite{goodell19frontiers,Pocher22,ballaschk21buba,Michalopoulos25}).
In this sense, auditability enables authorized parties to verify compliance properties or selectively access transaction information, rather than making transactions universally visible.
From a design perspective, auditability introduces a controlled relaxation of privacy guarantees. Increasing audit power typically strengthens trust assumptions or governance requirements and may reduce anonymity. Therefore, existing systems differ in what can be audited, when auditing may occur, and who is trusted to perform it.
Following a regulation-by-design approach, auditability mechanisms can be embedded directly into the protocol to reconcile compliance with privacy.
As discussed in Sections~\ref{sec:ppf:prot} and~\ref{sec:ppf:value}, the feasibility and scope of auditability mechanisms depend strongly on protocol architecture and on the underlying value representation.
As such, auditability represents an explicit design concern of third generation systems, rather than an isolated feature.

\paragraph{Auditability, Policy, and Accountability}
Auditability refers to the ability to examine, track, and verify transactions and balances while preserving a certain degree of privacy. 
It is a key feature of the third generation of privacy-preserving digital payment systems and, in particular, CBDCs.
Auditability operates between these notions by supporting \emph{post-facto} verification: it allows authorized parties to assess compliance or investigate suspicious behavior after transactions have occurred, and only within a prescribed scope.
Two related concepts clarify its role: policy and accountability. 
A \emph{policy} specifies admissible behavior and is typically enforced \emph{ex-ante} by preventing invalid transactions.
\emph{Accountability} concerns the ability to associate actions with responsible entities, often via identity-binding mechanisms established at registration time.

\paragraph{What is audited, in practice?}

Usually, auditability does not grant auditors unrestricted access to transaction data. Instead, it provides structured access to specific protocol state. 
This typically includes persistent state associated with users or tokens (e.g., counters or commitments), encrypted auxiliary metadata, and ZKPs attesting that policy constraints were respected.
Importantly, auditability is rarely tied to a single account or address. To prevent trivial circumvention (e.g., by generating fresh accounts), audit-relevant state is anchored to identity credentials, token histories, or cryptographically bound counters that persist across transactions.
As a result, address rotation does not reset spending history or policy enforcement.
Auditability mechanisms represent constraints on system evolution rather than direct inspection tools: they regulate how hidden state may evolve over time and enable authorized parties to verify compliance without pervasive disclosure.

\paragraph{Mechanisms}
We identify the following approaches that enable or support auditability: 
anonymity budget,
operating limits,
revocable anonymity and user tracing,
coin tracing,
and verification without disclosure.

\subsubsection{Anonymity Budget}
An \emph{anonymity budget} restricts how many anonymous payments a participant can perform, with each transaction consuming part of it (e.g.,~\cite{Camenisch06,wust19:prcash,tomescu22:utt,bank2019exploring}).
Camenisch et al.~\cite{Camenisch06} propose a \emph{bounded-anonymity business model}, where a public limit defines how many coins a user may anonymously transfer to a merchant.
PRCash~\cite{wust19:prcash} instead caps the total anonymous funds a user can \emph{receive} within a time window.
The anonymity budget is represented as a non-spendable output, whose size and duration regulate the flow of anonymous funds. Once exhausted---or if the user chooses non-anonymous transactions---the payee attaches his identity, encrypted under the regulator's key.
UTT~\cite{tomescu22:utt} applies the budget to \emph{spending}, requiring either loss of privacy or auditor intervention once the allocation is exceeded. Auditors can refresh budgets or approve extra transactions after requesting more details.
In all these systems, the budget is realized as a special token included in every anonymous yet accountable transaction.

\subsubsection{Operating Limits}
Operating limits indicate constraints on transactions and accounts, such as maximum value, user balance, total sent/received funds, or transaction count within a time window. 
Compliance is generally proven in zero-knowledge.
Garman et al.~\cite{Garman17} extend Zerocash with auditability features, including regulatory closure (i.e., restricting transactions to assets of the same type), spending limits, selective user inquiries, and coin tracing. A central idea is the \emph{per-transaction spending limit}: transactions above the limit are invalid unless signed by an authority. Another approach uses a \emph{per-user counter} incremented with each outgoing transaction. The system can then reject transactions beyond a threshold.
Gross et al.~\cite{gross2021} build on this with a two-layer account-based system for implementing a CBDC, where each account tracks \emph{epoch turnover} (total spent per epoch), proven with ZKPs.
Similarly, AQQUA~\cite{Papadoulis24:aqqua} maintains \emph{multiple counters} per account (balance, sent, received), publicly stored as three commitments. 
A registration authority links users identities to their (possibly, multiple) public keys, while an audit authority check compliance with five policy types: spending/receiving limits, transaction value limits, inactivity over a time period, selective transaction  disclosure.
During an audit, the authority selects a user by his public key and requests the opening of two historical account state snapshots (for every user's accounts).
After audits, the user randomizes his accounts (and provides a ZKP of valid update), thereby restoring his privacy.
Other systems also rely on counters.
Xue et al.~\cite{xue23} use per-user counters stored in a Merkle tree managed by regulators, which are also in charge of user on-boarding.
ZKPs\footnote{Specifically, zk-SNARKs are used for arithmetic statements and Sigma Protocols for algebraic statements.} on counters enable policies on total amount and \emph{frequency} of fund transfers within a time period.
When suspicious transactions are detected, the participants' identities can be recovered by regulators.
Chen et al.~\cite{chen20:pgc} define policies on transaction limits, rates, and value, also verified with ZKPs.
Across these proposals, auditability is achieved by maintaining cumulative state (e.g., counters or commitments) whose evolution is constrained by policy.

\subsubsection{Revocable Anonymity and User Tracing}
User tracing mechanisms enable authorities to discover a user's identity or link his transactions when required, implementing \emph{revocable} or \emph{conditional anonymity} models.\footnote{Revocable anonymity refers to a model where users remain anonymous by default, but their identity can be revealed under certain conditions. User tracing refers to technical mechanisms enabling authorized entities to identify users or link their transactions when required.}

Some schemes rely on special \emph{digital signatures} that expose the sender on double spending (e.g.,~\cite{hayes90,Zhang19,beer24:payoff}).
For example, Chaum-Fiat-Naor~\cite{chaum90} embed identifying information that is revealed upon double spending, while Schoenmakers~\cite{Schoenmakers1998} improves efficiency using ``once-concealed, twice-revealed'' proofs. Brands~\cite{Brands94} replaces blind signatures with credential-based encodings to support selective disclosure, and Camenisch-Lysyanskaya~\cite{CamenischLysyanskaya01} signatures further generalize these ideas to multi-attribute tokens with flexible revocation policies.

Other approaches embed \emph{encrypted} identifiers, keys, or metadata into \emph{tags}, accessible only to {tracing authorities}  (e.g.,~\cite{Cecchetti17,Garman17,Zhang19,liu22,lin20,li21:traceable-monero,Kiayias22:peredi}). This is often referred to as the \emph{trapdoor} approach.
Many systems also introduce a registration authority that links cryptographic keys to real-world identities (e.g.,~\cite{Zhang19,liu22}). 
For instance, Barki et al.~\cite{barki20} adopt this model in a permissioned setting, embedding encrypted identity information into transactions and enabling a designated revocation authority to deanonymize users when required.
Cecchetti et al.~\cite{Cecchetti17} use oblivious RAM to manage transactions across banks, while Liu et al.~\cite{liu22} encrypt user identities under the central bank's key.  
Zhang et al.~\cite{Zhang19} rely on a supervision authority with decryption power. 
Li et al.~\cite{li19net} propose an identity tag encrypted under a tracing authority's key.
Garman et al.~\cite{Garman17} propose a tracing mechanism where each user receives a \emph{unique encryption key}: if the user is being traced, the key is a randomized version of the authority's key; otherwise, it is derived from a null key.
Androulaki et al.~\cite{Androulaki20} encrypt transaction data under the auditors of both sender and receiver, while Xue et al.~\cite{xue23} scope tracing by encoding user identities with epoch identifiers.
The epoch identifier enables temporal scoping, which limits the impact of key compromising and can enable granular tracing\tlong{ (e.g., to investigate suspicious behavior in a particular time window)}.
Other systems derive \emph{one-time keys} from long-term keys, enabling tracing by recovering the latter.
Lin et al.~\cite{lin20} let users perform transactions using \emph{anonymous keys} derived from their public addresses. A manager can use his key and the anonymous key to recover the user's public address.
Traceable Monero~\cite{li21:traceable-monero} tags \emph{payee} one-time keys: an authority can decrypt the tag and perform \emph{backward and forward tracing} (i.e., identify the accounts that sent/receive funds to/from the current one).

A common concern is the trust concentration in a \emph{single tracing authority}.
To address this, Yu et al.~\cite{Yu24} split tracing between \emph{two authorities} via encrypted ``marks'' requiring joint computation, while PEReDi~\cite{Kiayias22:peredi} (and Sarencheh et al.~\cite{Sarencheh25:parscoin}) distribute decryption among multiple parties, requiring \emph{quorum} approval to identify users and access transaction metadata.
Similarly, PICTURE~\cite{Zhang24} adopts a threshold-based regulation model, where unlinkable transactions embed encrypted trace records that can be opened only by a quorum of authorities to recover sender and receiver identities.

\subsubsection{Coin Tracing}
Garman et al.~\cite{Garman17} enable tracing by marking specific \emph{tainted coins}, where all derived outputs remain traceable. Each coin carries a fresh \emph{encryption key}; tracing data for outputs is encrypted under both this key and the authority's public key, preventing unauthorized tracing by users. To avoid cascade tracing and collusion, exchanges can (verifiably) remove traceability with a public dummy key.
Keller et al.~\cite{keller21:collabdeanonym} study de-anonymization in CoinJoin and ring signatures through collaboration with \emph{witnesses}---i.e., users who confirm their inputs were not part of a mix. As the number of witnesses grows, the anonymity set shrinks, enabling more effective deanonymization.

\subsubsection{Verification without Disclosure}
Verification without disclosure lets verifiers confirm policy compliance without accessing private data. 
These mechanisms rely on cryptographic proofs (usually NIZK) to attest that hidden data satisfies constraints (e.g.,~\cite{Narula18,wust22:platypus,Cecchetti17,bunz20:zether,Ma21:dsc}).
The verifier learns only the truth of the statement. 
Besides the approaches in \emph{Operating limits}, here we mention those supporting policies beyond basic balance or transfer correctness check (e.g.,~\cite{bunz20:zether,miers13zerocoin,Fauzi19:quisquis,Ma21:dsc,baudet22:zef,Guan22:blockmaze,beer24:payoff,Papadoulis24:aqqua,friolo25}).
zkLedger~\cite{Narula18} uses Pedersen commitments 
enabling aggregate computations. 
\tlong{It employs generalized Schnorr-based NIZKs to prove the correctness of computations without revealing underlying data.}
zkLedger supports complex auditing via a map-reduce model: banks generate commitments over filtered and transformed data (e.g., checking for non-zero values or squaring for variance) along with NIZKs proofs of correctness, enabling auditors to compute metrics like means and variances.\footnote{Recently, Yang et al.~\cite{Yang25:zkfabledger} 
extend the zkLedger lineage to permissioned enterprise settings with auditable confidential asset transfers.}
Platypus~\cite{wust22:platypus} defines account states with opaque \emph{auxiliary data} to support custom policy checks. 
Solidus~\cite{Cecchetti17} uses PVORM and ZKPs 
to publicly verify a bank's transaction log compliance.
DSC~\cite{Ma21:dsc} enforces compliance of encrypted transactions with range and authorization proofs\tlong{, without
revealing transaction details}.
These approaches primarily support compliance verification rather than attribution: they enable auditors to assess whether rules were followed, without necessarily identifying the transacting parties.

\subsubsection{Discussion}
Choosing an auditability mechanism depends on the application context and threat model. Lightweight approaches such as anonymity budgets and operating limits are suitable for retail payments where limiting large-scale abuse is sufficient, but they offer limited forensic capabilities and are vulnerable to Sybil strategies unless combined with identity binding. Revocable anonymity provides stronger deterrence by enabling authorized identity recovery, at the cost of increased trust assumptions and governance complexity. Coin-tracing mechanisms support post-incident analysis of fund flows and recovery of tainted assets, but may weaken unlinkability and expose transaction graph structure. Verification without disclosure enables compliance checks (e.g., balance or limit enforcement) without revealing identities or values, but typically does not support attribution. Consequently, no single approach dominates the design space: compliance-oriented systems favor revocable or traceable designs, privacy-first systems rely on limited or aggregate auditability, and hybrid architectures increasingly combine multiple techniques to balance privacy, trust, and enforcement. \subsection{Summary and Takeaways}
\label{sec:ppf:summary}
This section highlights the evolving landscape of privacy goals in digital payment systems. 
Early approaches focused primarily on anonymity and unlinkability via address obfuscation and mixing, while recent designs emphasize confidentiality and auditability, enabled by advanced cryptographic tools such as ZKPs, homomorphic commitments, and threshold encryption.

Each privacy goal---anonymity, confidentiality, unlinkability, and auditability---relies on distinct mechanisms and entails trade-offs. For instance, unlinkability may hinder auditability, while encrypted transactions that preserve confidentiality can complicate oversight.
No single protocol satisfies all privacy and compliance objectives; systems tend to prioritize different goals based on their threat models, trust assumptions, and intended deployment contexts.
Table~\ref{tab:summary} summarizes how privacy goals are addressed in practice, mapping privacy properties to cryptographic primitives, system mechanisms, and representative protocols. This provides a reference  for researchers and practitioners navigating the trade-offs in privacy-preserving payment system design.

\begin{table*}[t]
\centering
{
\caption{Summary of privacy goals, main enabling technologies, and representative protocols}
\label{tab:summary}
{\footnotesize
\begin{tabular}{@{}p{2.2cm}p{4.6cm}p{4.6cm}p{3.1cm}@{}}
\toprule
\textbf{Privacy Goal} 
    & \textbf{Key Cryptographic Tools}
    & \textbf{System Mechanisms}
    & \textbf{Example Protocols} \\
\midrule
\textbf{Sender Anonymity} 
    & Blind, ring, group, and randomizable signatures, commitments and ZKPs
& One-time addresses, anonymity set, anonymous addresses
    & eCash, UTT, Zerocoin \\
\rowcolor{gray!10}
\textbf{Receiver Anonymity} 
    & One-time public keys, 
encrypted addresses, commitments
    & One-time addresses 
    & MimbleWimble, Quisquis, Zerocash \\
\textbf{Confidentiality} 
    & Commitments, encryption, ZKPs 
    & Confidential values and balances& Zether, PEReDi, Monero \\
\rowcolor{gray!10}
\textbf{Network Anonymity} 
    & TOR, I2P, Dandelion++ 
    & Obfuscate IP addresses and routing paths
    & Monero \\
\textbf{Unlinkability} 
    & Mixing protocols, ring signatures, one-time addresses, commitments 
    & Fresh key generation, decoy input, output set hiding 
    & Platypus, BlockMaze, CoinShuffle++ \\
\rowcolor{gray!10}
\textbf{Auditability} 
    & Authenticated tokens, threshold encryption, anonymous keys, trapdoor, ZKPs
    & Anonymity budget, operating limits, tracing hooks, revocable anonymity
    & PEReDi, Solidus, zkLedger \\
\bottomrule
\end{tabular}
}
}
\end{table*}
 
Privacy-preserving payment systems typically combine multiple cryptographic tools, each with distinct trade-offs. Blind signatures are lightweight but centralized; ring signatures provide scalable anonymity at latency cost; threshold encryption and commitments support auditability but not anonymity; and ZKPs (e.g., zkSNARKs, zkSTARKs) offer strong guarantees, but they are computationally intensive, may require substantial RAM (e.g.,~\cite{Almashaqbeh22,liang25:soksnarks}), and require careful engineering to achieve practical deployment.
One of the well-known ZKP-based systems is zk-SNARK, which makes transactions compact for verification but needs a trusted setup that, if mishandled, could compromise the security of the system.\footnote{In practice, parameters are generated via publicly auditable multi-party computation ceremonies, and security holds as long as at least one participant is honest and destroys the corresponding secret randomness.} 
Importantly, as shown in Section~\ref{sec:ppf:audit}, implementing auditability often requires careful design of ledger data structures and interaction protocols to avoid unintended information leaks while still enabling policy verification. 
This multidimensional design space highlights the need for a holistic, design-oriented perspective, where privacy-preserving digital payment systems combine complementary mechanisms to balance user privacy with institutional accountability.
 \section{Open Challenges and Research Directions}
\label{sec:openchallenges}

Despite significant advances, numerous open challenges remain at the intersection of business requirements, system design, and expectations of privacy and auditability. 
We identify a few interesting research directions.

\subsection{Formal Definitions of Privacy and Auditability}
Existing systems use ad-hoc definitions for privacy and auditability, and previous works (e.g.,~\cite{genkin18,Almashaqbeh22}) highlight theoretical gaps in unified privacy and auditability models. 
Most protocols informally claim anonymity, unlinkability, or confidentiality, but lack formal, composable definitions covering graph obfuscation, timing metadata, value-hiding, selective disclosure, and conditional anonymity.
Systems combining multiple primitives rarely have UC~\cite{Canetti00:UC} or IITM~\cite{Kuesters13:IITM} proofs, and auditability features (e.g., view keys, trapdoors) are seldom formalized.

\subsection{Scalability and Efficiency}
Many cryptographic protocols and PETs remain computationally demanding (e.g., homomorphic encryption, ZKPs, and MPC).
Future research must focus on developing optimized, lightweight, and parallelizable privacy-preserving protocols that can operate efficiently in high-throughput environments. 
Promising directions include hardware acceleration (e.g., via TEEs) and the exploration of protocol variants that balance privacy guarantees with resource efficiency.
ZKPs play a central role in many systems, yet general-purpose schemes (e.g., zk-SNARKs, zk-STARKs) still face challenges related to trusted setup, proof aggregation, and post-quantum security.
For instance, zk-SNARKs typically require a trusted setup, introducing centralization risks and limiting composability. While transparent SNARKs (e.g., Plonky2\footnote{\url{https://github.com/0xPolygonZero/plonky2}} and Spartan~\cite{Setty19:spartan}) and STARKs mitigate these concerns, they often incur higher computational costs. Achieving efficient, setup-free ZKPs without compromising performance remains an open research problem.
Additionally, support for efficiently generating proofs that can verify other proofs---known as recursive proof composition---is still limited, even though this capability is crucial for scalable applications like off-chain rollups. Developing ZKPs that are not only succinct and recursive, but also universally verifiable (i.e., verifiable by anyone without special access), remains a technically challenging and active area of research.
No current ZKP scheme offers simultaneously strong post-quantum guarantees\footnote{ZKPs based on bilinear pairings (e.g., Plonk, Groth) or Pedersen vector commitments (i.e., Bulletproof) assume at least the discrete logarithm problem, hence are subject to quantum threats. Other SNARKs (e.g., Ligero, Orion) and STARKs using Merkle tree and other protocols to build a polynomial commitment are post quantum resistant.} and performance comparable to elliptic curve-based constructions.

\subsection{Balancing Privacy with Regulatory Compliance}
Total anonymity is incompatible with regulatory oversight, making auditable privacy and selective disclosure critical design goals.
An open challenge is how to enable systems where honest users maintain strong privacy guarantees, while still allowing authorities to investigate illicit activities under well-defined conditions. 
While preliminary approaches exist, significant work remains to formalize 
disclosure policies, define the types of information subject to auditing, and clarify the roles and interactions of participating entities.
Relevant research directions include ZKPs with selective opening, trapdoor, witness encryption, and anonymity budgeting---each enabling varying levels of conditional transparency.
Less explored primitives, such as witness encryption, 
may enable advanced privacy-preserving policies
(see \refAppWitnessEnc{}).
Additionally, emerging requirements, such as time-bound anonymity, jurisdiction-aware confidentiality, and programmable disclosure, call 
for flexible and context-aware privacy controls.
Designing PETs that can support such adaptive 
behaviors, while remaining robust against misuse, poses a complex challenge involving technical, legal, and usability considerations.

\subsection{Adaptive Privacy and Role Rotation}

Privacy-preserving systems rely on anonymity sets, yet managing 
cryptographically sound methods for dynamic and attack-resistant management remain lacking, especially in decentralized or evolving environments.
Current schemes, such as fixed-size ring signatures or heuristic decoy selection, degrade as usage patterns shift and are vulnerable to poisoning attacks, where adversaries insert tainted outputs to weaken privacy. 
Beyond anonymity sets, adaptive privacy must address the evolving nature of system participants, especially in regulatory or issuance roles. In CBDCs and regulated payment systems, auditors, compliance entities, or issuers may change due to policy or organizational shifts. Supporting this evolution without compromising anonymity or confidentiality is difficult: most systems assume static trust anchors or fixed multi-signature sets. Cryptographic tools such as threshold cryptography, distributed key generation, or policy-based encryption could enable dynamic reconfiguration, but balancing adaptability with unlinkability, correctness, and forward secrecy remains unresolved.

\subsection{Network-Level Anonymity and Cross-layer Privacy Leakage}

Network-layer anonymity 
is typically outsourced to external systems (e.g., Tor, mix networks).
However, native integration of network-level privacy mechanisms into privacy-preserving payment system protocols remains largely unresolved.
This delegation creates a critical cross-layer vulnerability, where even perfectly private on-chain transactions can be de-anonymized through timing analysis, transaction ordering, volume inference, or network topology correlation.
Existing protocols often assume independence between transaction content and communication patterns, which adversaries can exploit. For instance, gossip-based mempool propagation leaks significant metadata about the origin and timing of transactions, enabling traffic correlation attacks.
Emerging research directions include the design of privacy-preserving mempools that incorporate indistinguishable broadcasting, cover traffic, oblivious relaying, or dandelion-style~\cite{fanti18:dandelionplusplus} diffusion to mask transaction origins. These approaches aim to reduce the information available to global or local adversaries observing network traffic (e.g.,~\cite{diaz2021nym,bebel22:ferveo}). However, these solutions often introduce latency, bandwidth overhead, or synchronization issues, leading to difficult trade-offs between efficiency, liveness, and anonymity guarantees.
Addressing network anonymity in digital payment systems requires cross-layer designs that align messaging protocols with the cryptographic guarantees of the ledger. 
This remains a highly active research area in adversarial environments. \section{Conclusion}
\label{sec:conclusion}

Designing digital payment systems that balance user privacy with institutional auditability remains one of the most intricate and impactful challenges in modern cryptographic and systems engineering.
Over the past decade, crypto-asset systems and CBDC prototypes have pushed innovation across cryptographic protocols, PETs, and system architectures.
Yet, despite substantial progress, the design space remains fragmented, and no universally accepted blueprint has emerged.

This survey provides a structured and technically grounded overview of the evolving landscape of privacy-preserving digital payment systems.
We systematize the literature across 
both decentralized and centrally managed designs, including crypto-asset systems and CBDC proposals, 
introducing a comprehensive taxonomy that connects privacy and auditability goals with underlying cryptographic primitives, protocol designs, and system architectures. 
In doing so, we bridge conceptual objectives (such as anonymity, confidentiality, and auditability) with concrete technical mechanisms and design models.

Our analysis traces the evolution of privacy-preserving digital payment systems across three generations, reflecting an increasing interest in addressing privacy-accountability trade-offs. We underscore that privacy is not a monolithic property but a multifaceted design goal shaped by system requirements and regulatory constraints.
Finally, we outline several open research challenges that call for interdisciplinary collaboration across cryptography, distributed systems, economics, and policy. Key directions include the development of scalable and composable privacy primitives, architectures that balance auditability with confidentiality, and formal models that capture realistic adversarial and compliance scenarios.
As digital payment technologies progress from academic prototypes to deployed systems, balancing strong privacy guarantees with regulatory compliance will require ongoing technical advances and interdisciplinary efforts.

\begin{acks}
We thank Marco Benedetti, Claudia Biancotti, Giuseppe Galano, Sara Giammusso, Antonio Muci, and Fabiana Rossi 
for their insightful discussions in the early stages of this work and for their valuable feedback throughout its development.
\end{acks}

\bibliographystyle{ACM-Reference-Format}
\bibliography{biblio}

\ifshowappendix
\newpage

\appendix
\fancyfoot[RO,LE]{}
\setcounter{page}{1}
\renewcommand{\shorttitle}{Supplementary Material}
\renewcommand{\shortauthors}{Nardelli et al.}

{
\noindent{\Large\vspace{0.5em}\textsc{Supplementary Material to:}\\ 
\LARGE\sffamily\bfseries
\thetitle{}
}
}

\vspace{1em}

{\noindent {\large\sffamily \uppercase{Matteo Nardelli, 
Francesco De Sclavis, 
Michela Iezzi},}
{\small\normalfont Bank of Italy, Italy
}}
\vspace{1.3em}

\section{Appendix}
This appendix provides supplementary material for \emph{\thetitle{}}, which complements the main analysis but is excluded from the core sections to preserve narrative flow. 
Specifically, it includes:
\emph{(i)} the research methodology used for the literature review (Section~\ref{apx:meth}); 
\emph{(ii)} a breakdown of publication venues for the works reviewed in this survey (Section~\ref{apx:sec:venue}), and
\emph{(iii)} an analysis of mixing protocols (Section~\ref{apx:mixing}) that complements the analysis in Section~\ref{sec:ppfeatures}, with particular emphasis on unlinkability techniques discussed in Section~\ref{ppf:aud:unlinkability}.
These materials are presented as tables and offer additional context and technical detail for readers interested in the broader research landscape and protocol-level design trade-offs.
Furthermore, in Section~\ref{app:emergincryptotools}, we briefly present interesting yet underexplored cryptographic tools that can be used to design privacy-preserving digital payment systems.

\subsection{Selection and Classification of the Reviewed Contributions}
\label{apx:meth}
To identify the relevant contributions we analyzed in the paper, we adopted the following approach: 
\begin{enumerate}
    \item We identified a core collection of works relying on direct authors' knowledge of the research field;
    \item We enlarged this collection through various iterations of \emph{backward snowballing}~\citeapndx{Wohlin14:systlitrev} (i.e., using the references), which allowed us to identify related works not included in the first step;
    \item Analyzing the selected papers, we identified the most frequent publication venues (i.e., journals or conferences with 2+ selected papers). Checking the list of papers published in these venues, we added further papers to the collection; 
    \item As a final step, we performed keyword-based searches on publication databases (i.e., Google Scholar, DBLP, Elsevier Scopus) using combinations of terms such as privacy, auditability/accountability, cryptocurrency/blockchain, crypto-assets, payment systems/CBDC, 
    and specific mechanism keywords (e.g., zero-knowledge proofs, anonymous/confidential transactions, conditional/revocable anonymity, tracing/AML compliance).
\end{enumerate}
After searching and scraping, we selected and reviewed about $70$ publications for this survey.
Section~\ref{apx:sec:venue} provides information about the most popular publication venues. Table~\ref{tab:sota-cryptocurrencies} summarizes the analysis of work related to privacy preserving digital payment systems and crypto-asset systems, whereas Table~\ref{tab:sota-mixing} reports the analyzed mixing protocols.

\subsection{Most Popular Publication Venues}
\label{apx:sec:venue}
Table~\ref{tab:venues} summarizes the most common publication venues among the reviewed works.
We report only the most frequent venues (with $ 2+ $ works), including journals (e.g., IEEE Transactions on Dependable and Secure Computing, Future Generation Computer Systems, IEEE Transactions on Information Forensics and Security), international conferences (e.g., Financial Cryptography and Data Security (FC), ACM Conference on Computer and Communications Security (CCS), IEEE Symposium on Security and Privacy (S\&P)), and other (non-peer reviewed) sources, such as pre-prints and whitepapers.

\begin{table}[h] \centering
\caption{Most popular publication venues among the reviewed works.}
\label{tab:venues}
\resizebox{.78\textwidth}{!}{
\begin{tabular}{lrl}
\toprule
\textbf{Venue Type} & \textbf{Number of Publications} & \textbf{Frequent Venues}  ($\geq$ 2 works)\\
\midrule
Journal & 16 & 
{
\begin{tabular}{l}
IEEE Trans. Dependable Secur. Comput. (6)\\
Fut. Gen. Comp. Sys.	(2) \\ 
IEEE Trans. Inf. Forensics Secur. (2)
\end{tabular}
}
\\
\midrule
Conference & 33 & 
{
\begin{tabular}{l}
FC	     (8)\\
ACM CCS  (5)\\ ESORICS	(3)\\
IEEE S\&P (3)\\
AFT	(2)\\
ASIACRYPT	(2)\\
CRYPTO	(2)\\
\end{tabular}
}\\
\midrule
Other & 16 & 
{
\begin{tabular}{l}
IACR Cryptol. ePrint Arch.	(8)\\
Whitepapers (7)
\end{tabular}
}\\
\bottomrule
\end{tabular}
}
\end{table} 

\subsection{Analysis of Mixing Protocols}
\label{apx:mixing}
Similar to Table~\ref{tab:sota-cryptocurrencies}, which surveys privacy-preserving digital payment and crypto-assets systems, Table~\ref{tab:sota-mixing} focuses specifically on mixing protocols. The table serves as a reference to facilitate comparison across existing approaches in the literature.
The corresponding symbols in Table~\ref{tab:sota_mixing_legend} extend those introduced in Table~\ref{tab:sota_legend}.

\renewcommand{\arraystretch}{0.9}
\setlength{\tabcolsep}{3pt}
{
\footnotesize
\begin{longtable}{lrlrl}
\toprule
\textbf{Group} & \multicolumn{4}{l}{\textbf{Legend}}\\
\midrule
\multicolumn{5}{c}{\textit{(continued from previous page)}}\\
\endhead
\caption{Additional labels used in Table~\ref{tab:sota-mixing} beyond those defined in Table~\ref{tab:sota_legend}.}
\label{tab:sota_mixing_legend}\\
\toprule
\textbf{Group} & \multicolumn{4}{l}{\textbf{Legend}}\\
\midrule
\endfirsthead
\midrule
\multicolumn{5}{c}{\textit{(continued on next page)}}\\\endfoot
\endlastfoot
\multicolumn{3}{l}{\emph{Protocol}}\\
{No-Trust Assumption} & \CIRCLE & Accountable & & \\
\midrule
\multicolumn{3}{l}{\emph{Other}}\\
{Cryptographic} & DC-net & Dining cryptographers network & HDC-net & Harmonized DC-net \\
{$\hookrightarrow$ Primitives} & MPC & Multi-party computation & TEE & Trusted Execution Environment \\
{} & VS & Verifiable shuffling & & \\
\bottomrule
\end{longtable}
\renewcommand{\arraystretch}{1.0}
}
 {
\renewcommand{\arraystretch}{0.9}
\setlength{\tabcolsep}{3pt}
\footnotesize
\begin{longtable}{lccccccccclc}
\caption{Classification of existing mixing protocols (see notation in Table~\ref{tab:sota_mixing_legend}).}
\label{tab:sota-mixing} \\
\toprule
\rowcolor{white}
\multicolumn{1}{l}{\textit{}} 
        & \multicolumn{6}{c}{\textit{Protocol}} 
        & \multicolumn{4}{c}{\textit{Privacy}} 
        & \multicolumn{1}{c}{\textit{Other}} 
\\
\cmidrule(r){2-7}
\cmidrule(r){8-11}
\cmidrule(r){12-12}
\rowcolor{white}
\textbf{Paper} 
        & \rv{Year} 
            & \rv{Decentralized} 
            & \rv{Blockchain Compatibility} 
            & \rv{No-Trust Assumptions} 
            & \rv{Implemented}
            & \rv{Different Denominations}
            & \rv{Sender Anonymity} 
            & \rv{Receiver Anonymity}
            & \rv{Transaction Value Confidentiality}
            & \rv{Unlinkability} 
        & \rv{Cryptographic Primitives} 
    \\
\midrule\endhead
\multicolumn{12}{c}{\textit{(continued on next page)}}\\\endfoot
\endlastfoot
CoinJoins~\cite{Maxwell13:coinjoin} & 2013 & \checkmark & \checkmark & \checkmark & $\times$ & $\times$ & $\times$ & $\times$ & $\times$ & \LEFTcircle & DS  \\ 
\rowcolor{gray!10}
CoinShuffle~\citeapndx{ruffing14coinshuffle} & 2014 & \checkmark & \checkmark~~(B) & \checkmark & $\times$ & $\times$ & $\times$ & $\times$ & $\times$ & \checkmark & DS, PE  \\ 
Mixcoin~\cite{Bonneau14:mixcoin} & 2014 & $\times$ & \checkmark~~(B) & \CIRCLE & $\times$ & $\times$ & $\times$ & $\times$ & $\times$ & \checkmark & DS  \\ 
\rowcolor{gray!10}
Xim~\cite{Bissias14:xim} & 2014 & \checkmark & \checkmark~~(B) & \checkmark & $\times$ & \checkmark & $\times$ & $\times$ & $\times$ & \checkmark & PE  \\ 
Blindcoin~\cite{Valenta15:blindcoin} & 2015 & $\times$ & \checkmark~~(B) & \CIRCLE & $\times$ & $\times$ & $\times$ & $\times$ & $\times$ & \checkmark & BS  \\ 
\rowcolor{gray!10}
CoinShuffle++~\cite{ruffing16:coinshuffleplusplus} & 2016 & \checkmark & \checkmark~~(B) & \checkmark & \checkmark\footnote{\url{https://github.com/decred/cspp}}& $\times$ & \checkmark & \checkmark & $\times$ & \checkmark & DC-net  \\ 
TumbleBit~\citeapndx{Heilman16:tumblebit} & 2016 & $\times$ & \checkmark~~(B) & \checkmark & $\times$ & \checkmark & \checkmark & \checkmark & $\times$ & \checkmark & MPC, ZK  \\ 
\rowcolor{gray!10}
SecureCoin~\citeapndx{ibrahim17:securecoin} & 2017 & \checkmark & \checkmark~~(B) & $\times$ & \checkmark & $\times$ & $\times$ & \checkmark & $\times$ & \checkmark & SA  \\ 
ValueShuffle~\cite{ruffing17valueshuffle} & 2017 & \checkmark & \checkmark~~(B) & \checkmark & \checkmark & \checkmark & \checkmark & \checkmark & \checkmark & \checkmark & PE, SA  \\ 
\rowcolor{gray!10}
CoinParty~\cite{Ziegeldorf18coinparty} & 2018 & \checkmark & \checkmark~~(B) & \checkmark & $\times$ & $\times$ & $\times$ & $\times$ & $\times$ & \checkmark & DS$^t$, Mixnets, MPC  \\ 
Liu et al.~\citeapndx{liu18mixing} & 2018 & $\times$ & \checkmark & \checkmark & $\times$ & $\times$ & $\times$ & $\times$ & $\times$ & \checkmark & RS  \\ 
\rowcolor{gray!10}
M\"{o}bius~\citeapndx{meiklejohn18:mobius} & 2018 & \checkmark & \checkmark~~(E) & \checkmark & \checkmark & $\times$ & \checkmark & \checkmark & $\times$ & \checkmark & RS,SA  \\ 
Obscuro~\citeapndx{tran18:obscuro} & 2018 & $\times$ & \checkmark~~(B) & \checkmark & \checkmark\footnote{\url{https://github.com/BitObscuro/Obscuro}}& $\times$ & $\times$ & $\times$ & \checkmark & \checkmark & TEE  \\ 
\rowcolor{gray!10}
MixEth~\cite{seres19:mixeth} & 2019 & \checkmark & \checkmark~~(E) & \checkmark & \checkmark\footnote{\url{https://github.com/seresistvanandras/MixEth}}& $\times$ & $\times$ & $\times$ & $\times$ & \checkmark & VS  \\ 
AMR~\citeapndx{le21:amr} & 2021 & \checkmark & \checkmark~~(E) & $\times$ & \checkmark & $\times$ & $\times$ & $\times$ & $\times$ & \checkmark & ZK  \\ 
\rowcolor{gray!10}
BlindHub~\cite{qin23:blindhub} & 2023 & $\times$ & \checkmark~~(B) & \checkmark & \checkmark\footnote{\url{https://github.com/blind-channel/blind-hub}}& \checkmark & $\times$ & $\times$ & $\times$ & \checkmark & BS*  \\ 
UCoin~\cite{Nosouhi23:ucoin} & 2023 & \checkmark & \checkmark~~(B) & \checkmark & $\times$ & $\times$ & $\times$ & $\times$ & $\times$ & \checkmark & HDC-net  \\ 
\rowcolor{gray!10}
Wang et al.~\cite{Wang23} & 2023 & \checkmark & \checkmark & \checkmark & \checkmark & $\times$ & $\times$ & $\times$ & $\times$ & \checkmark & MT  \\ 
\end{longtable}
\renewcommand{\arraystretch}{1.0}
}
 
\subsubsection{Protocol}

\paragraph{System Architecture}
Both centralized (e.g.,~\cite{Bonneau14:mixcoin,Valenta15:blindcoin,qin23:blindhub} and \citeapndx{Heilman16:tumblebit,liu18mixing}) and decentralized solutions (e.g.,~\cite{Ziegeldorf18coinparty,Bissias14:xim,ruffing17valueshuffle,Nosouhi23:ucoin,Wang23} and~\citeapndx{ruffing14coinshuffle}) have been proposed to implement mixing protocols.
In centralized solutions, a third party (\emph{mixer} or \emph{tumbler}) provides the mixing service for a fee.
While such designs are simple and support large anonymity sets, they introduce risks such as availability, theft, and potential privacy breaches. 
Therefore, different trust models have been proposed to mitigate these concerns (see Section~\ref{sec:ppf:trust}).
Decentralized mixers remove the need for a trusted party by using cryptographic protocols or peer-to-peer coordination, enhancing security and censorship resistance, but they often incur higher complexity, require user coordination, and typically support smaller anonymity sets (e.g., Xim~\cite{Bissias14:xim}, CoinShuffle++~\cite{ruffing16:coinshuffleplusplus}, and  ValueShuffle~\cite{ruffing17valueshuffle}).
Both centralized and decentralized approaches face challenges such as Sybil attacks\footnote{In a Sybil attack, an adversary injects fake participants into the mix to degrade anonymity.} and regulatory scrutiny.

\paragraph{Trust Assumptions}
As regards centralized mixing protocols, 
since the third party represents a single point of failure, different protocols focused on different threat models,
where the mixer may be \emph{trusted and accountable} (e.g.,~\cite{Bonneau14:mixcoin,Valenta15:blindcoin}) or \emph{untrusted} (e.g.,~\citeapndx{Heilman16:tumblebit}). 
For instance, Mixcoin offers a signed warranty that will enable his customers to unambiguously prove if the mix has misbehaved~\cite{Bonneau14:mixcoin}. 
Conversely, Tumblebit~\citeapndx{Heilman16:tumblebit} prevents coin thief using an on-chain escrow and ZKPs.
The rise of off-chain solutions has let to the exploration of alternative mixer definitions. 
For instance, BlindHub~\cite{qin23:blindhub} implements mixing within payment channel hubs, where the hub acts as the tumbler. It employs blind signature-based protocols to hide users' addresses from the hub, ensuring privacy.

\subsubsection{Privacy Features}
\label{apx:sec:mix:privacy}

\paragraph{Unlinkability}
Different mixing protocols have been proposed (see \refTabMixing{}).
The na\"{i}ve approach for mixing transactions consists in a centralized service that collects transactions, and publishes on the ledger a single large transaction with multiple inputs and outputs. This hides the information regarding which inputs flows towards which output, thus limiting (if not preventing) linkability between sender and receiver. 
The Mixcoin centralized protocol considers an accountable mixer{, which offers a signed warranty that will enable his customers to unambiguously prove if the mix has misbehaved}~\cite{Bonneau14:mixcoin}. Blindcoin~\cite{Valenta15:blindcoin} extends this protocol by considering blind signatures for the warrants; this hides the link between input and output addresses of a user even from the mixer, achieving full unlinkability.
Tumblebit~\citeapndx{Heilman16:tumblebit} prevents coin thief using an on-chain escrow and ZKPs.
In BlindHub~\cite{qin23:blindhub}, mixing is performed within payment channel hubs, where the hub acts as tumbler and uses blind signatures to hide the customers' addresses from itself.
Obscuro~\citeapndx{tran18:obscuro} is a mixing protocol that utilizes Trusted Execution Environments (TEEs) and enables participants to verify the mixer's integrity via remote attestation.\footnote{Obscuso is specifically designed to defend against two threats: the participation rejection attack, where malicious users try to manipulating mixing by withdrawing their deposits before the TEE reads them, and the blockchain forking attack, where adversaries present stale blocks to mislead the mixer.}
Numerous papers also investigate decentralized protocols. 
CoinJoin~\cite{Maxwell13:coinjoin} is a technique that allows multiple users to combine their transactions into a single joint transaction; however, the author only describes the mixing idea. CoinShuffle~\citeapndx{ruffing14coinshuffle} allow to combine multiple users’ inputs and outputs in a single transaction but may compromise anonymity if transaction values differ or the anonymity set is small. 
Later, Ruffing et al.~\cite{ruffing16:coinshuffleplusplus} proposed CoinShuffle++, which uses a more efficient mechanism for anonymity, i.e., \emph{Dining Cryptographers Networks} (DC-nets) instead of mix-nets: while a mix-net requires sequential processing, hence a number of communication rounds linear in the number of participants, DC-nets enable to process mixing in parallel, hence it requires only a constant number of communication rounds.
Xim~\cite{Bissias14:xim} is a multi-round protocol for anonymously finding mix partners based on advertisements placed on the blockchain, with the property that no outside party can identify participants that pair up. 
SecureCoin~\citeapndx{ibrahim17:securecoin} creates an aggregated temporary deposit address and requires the collaboration of involved peers to generate novel destination addresses and collectively redistribute coins.
Leveraging \emph{Confidential Transactions} (which nonetheless is available only in a soft fork of Bitcoin), ValueShuffle~\cite{ruffing17valueshuffle} provides full privacy besides parties anonymity.
Confidentiality enables mixing funds of different values without compromising unlinkability.
CoinParty~\cite{Ziegeldorf18coinparty}  is a decentralized mixing protocol that distributes the mixing process across a group of semi-trusted parties\tlong{, called mixing peers}. They use an MPC protocol to jointly perform sensitive operations (i.e., encryption, decryption, and address shuffling), ensuring that no single party can link input and output addresses. \tlong{Each participant's output address is encrypted and shuffled multiple times by different peers. }Threshold signatures is used to enable claiming funds upon agreement of a majority of mixing nodes\tlong{ (thus improving protocol robustness)}. It also includes mechanisms to detect and exclude misbehaving participants, increasing the overall protocol reliability.

Decentralized tumblers have also been designed using \emph{smart contracts} on Ethereum.
M\"{o}bius~\citeapndx{meiklejohn18:mobius} uses ring signatures and stealth addresses to obfuscate the senders and recipients associations. First, senders furnish both their funds and stealth keys to the smart contract; then, each recipient creates a ring signature to withdraw their funds from the contract and transfer them to an ephemeral address. The size of the anonymity set is limited to the size of the ring, and the gas cost of the withdrawing transaction increases linearly with the size of the ring. 
MixEth~\cite{seres19:mixeth} aims to minimize gas costs by utilizing Neff's verifiable shuffle~\citeapndx{neff01:verifiableshuffle} and off-chain messages. Once the sender deposits coins, the MixEth contract enters rounds of shuffling and challenging, during which anyone can challenge the correctness of the preceding shuffle. When recipients believe that sufficient shuffling has been performed, they can withdraw their funds from the MixEth contract. 
\tlong{Although the shuffling occurs off-chain, MixEth stores the newly shuffled public keys and all the necessary information within the contract to enable anyone to verify the shuffle's correctness and continue public key shuffling after the corresponding challenge round.}We also mention AMR~\citeapndx{le21:amr} which disrupts the traceability connection between coins deposited and withdrawn by the same user. Participants deposit a predetermined quantity of coins into a smart contract, which establishes a Merkle tree structure over the deposits. 
To withdraw coins, a zk-SNARK 
prove coins inclusion 
in the Merkle tree. AMR leverages lending platforms to provide users with the opportunity to accrue interest on their deposited assets.

Mixing protocols are still vulnerable to transaction graph analysis to some extent.
For example, if the mixing process does not involve adequate randomization or if there are patterns in the amounts or timing of transactions, likely connections can be identified, even after several rounds of mixing. 
To maximize privacy, participants need to combine mixing protocols with other privacy-preserving techniques \tlong{(e.g., stealth addresses, payment channels) }and follow best practices (e.g., distributing coins across multiple addresses) to 
reduce the risk of being traced through 
graph analysis.
Further details on mixing solutions for Bitcoin and Ethereum, e.g., in~\cite{arbabi23:mixingsurvey}, and on transaction graph analysis, e.g., in~\cite{ron13:txgranalysis,chan17icitst} and in~\citeapndx{fleder15}.

\subsection{Emerging Cryptographic Primitives}
\label{app:emergincryptotools}

In addition to the cryptographic tools discussed in the main survey, several advanced primitives offer promising tools for future research in privacy-preserving digital payment systems.
Although not yet widely adopted in practical systems, these techniques may enable more flexible or robust privacy features in regulated or decentralized contexts.

\subsubsection{Anonymous Credentials}
Systems like Idemix~\citeapndx{Camenisch02:idemix} and Coconut~\cite{sonnino20:coconut} allow users to prove possession of attributes (e.g., age, residency, KYC compliance) without revealing their identity.
These systems support selective disclosure, enabling users to present only the minimal information required for a given interaction; 
for example, proving age above $18$ without disclosing birthdate.
This selective transparency could be very important for digital payment systems subject to regulatory constraints (e.g., AML, KYC), 
as users can demonstrate compliance without full deanonymization.
Moreover, advanced anonymous credential schemes could support unlinkability, preventing different credential uses from being correlated.
However, at the time of writing, efficiently revoking credentials (e.g., after misuse or expiration) while preserving privacy is nontrivial and typically requires accumulator-based techniques or dynamic revocation protocols. 
As digital payments increasingly intersect with real-world identity frameworks, anonymous credentials offer a promising path for balancing privacy and regulatory needs.

\subsubsection{Cryptographic Accumulators} 
Accumulators provide compact representations of sets along with efficient membership proofs. They are useful for constructing privacy-preserving mechanisms such as spent coin tracking, revocation, and anonymous group membership verification.
Although some systems already use Merkle trees to prove coin or user inclusion in a list of valid or authorized entities (e.g.,~\cite{Sander99,gross2021,xue23,Wang23,elfadul24}), we believe that accumulators remain underutilized—particularly in the design of revocation mechanisms, where they could offer efficient and privacy-preserving solutions.

\subsubsection{Functional Encryption}
Functional encryption~\citeapndx{Boneh10funcenc} allows specific functions to be computed on encrypted data without revealing the underlying inputs. 
In contrast to traditional public-key encryption, where decryption reveals the entire message, functional encryption supports function-restricted decryption keys.
In the context of digital payment systems, this could enable selective disclosure, e.g., allowing regulators to verify policy compliance or transaction conditions without full access to transaction details.
For instance, a regulator could decrypt only whether a transaction exceeds a policy-defined threshold without learning the actual value.
Potential applications may include: 
\emph{(i)} allowing a central bank to verify that a transaction is below a predefined limit, without revealing the amount;
\emph{(ii)} enabling eligibility verification for financial services based on encrypted KYC attributes;
and \emph{(iii)} constructing privacy-preserving tax collection mechanisms that expose only aggregate (or policy-relevant) outcomes.
Despite its promise, functional encryption remains largely theoretical for high-complexity functions and faces efficiency and key management challenges in practical deployments.

\subsubsection{Time-Lock Encryption}
Time-lock encryption~\citeapndx{Liu18timelockenc} delays access to encrypted information until a predefined time has elapsed.
Potential applications include time-delayed auditability, sealed transactions, or rate-limited disclosure of account (sensitive) data.

\subsubsection{Witness Encryption}
\label{app:witenc}
In its canonical form, witness encryption~\citeapndx{Garg13witness} allows one to encrypt a message with respect to an instance $ x $ of a (nondeterministic polynomial time $NP$) language, $L\in NP$, such that decryption is possible if and only if the decryptor possesses a valid witness $w$ for the statement $x\in L$. This capability opens the door to a number of expressive access control mechanisms that could be beneficial in the context of digital payment systems.
One promising application is \emph{signature-based witness encryption}, where the $NP$ statement encodes a threshold condition over a set of public keys. For example, the instance $x$ could describe a set of $n$ public keys, and the corresponding witness would be any subset of $k$ valid signatures under distinct keys.
This effectively enables a form of threshold encryption---where decryption requires authorization from at least $k$ participants---without any setup phase for correlated key generation or key sharing, unlike traditional threshold crypto-systems.
While a downside is that ciphertext size can grow with the complexity of the underlying relation (e.g., linear in $n$), recent constructions have demonstrated ciphertexts of size $O(\log n)$,
under specific assumptions (i.e., indistinguishability obfuscation for Turing machines),
making this approach increasingly practical~\citeapndx{avitabile24swe}.
Another line of research could explore \emph{witness encryption with hidden statements}, where the encrypted message can only be decrypted by someone knowing a witness for a secret statement, while the statement itself remains concealed. 
This property could offer a powerful tool for regulatory compliance. 
For example, policy enforcement could be realized by requiring users to provide ZKPs of compliance (i.e., knowledge of a valid witness), without revealing the rule being enforced.
Such a mechanism may be appealing to regulators who wish to enforce certain constraints (e.g., AML/KYC policies, tracing) without disclosing the exact criteria 
(e.g.,~\citeapndx{mu24:instancehidingip}).
To this end, users must include a secret witness with each transaction, attesting to their compliance with the policy.  
To conclude, witness encryption could provides a rich 
foundation for designing advanced privacy and access control features,
especially in contexts where minimal trust assumptions, fine-grained control, and privacy coexist.

These emerging primitives highlight exciting directions for foundational research in the design of privacy-preserving digital payment solutions. 
While most remain experimental, their potential to address regulatory, architectural, or usability gaps makes them worth further exploration.
  \bibliographystyleapndx{ACM-Reference-Format}
\bibliographyapndx{biblio}
\fi

\end{document}